\documentclass[11pt, dina4,  amssymb]{article}
\usepackage{amssymb, amsmath,amsthm}
\usepackage{graphicx}
\usepackage{multirow}
\usepackage{epstopdf}
\usepackage{color}
\usepackage[margin=1in]{geometry}

\newcommand{\p}{\partial}

\newcommand{\Og}{\Omega}
\newcommand{\fl}[2]{\frac{#1}{#2}}

\newcommand{\gm}{\gamma}
\newcommand{\nn}{\nonumber}
\newcommand{\ap}{\alpha}
\newcommand{\bt}{\beta}

\newcommand{\Dt}{\Delta}

\newcommand{\be}{\begin{equation}}
\newcommand{\ee}{\end{equation}}
\newcommand{\ba}{\begin{array}}
\newcommand{\ea}{\end{array}}
\newcommand{\bea}{\begin{eqnarray}}
\newcommand{\eea}{\end{eqnarray}}
\newcommand{\beas}{\begin{eqnarray*}}
\newcommand{\eeas}{\end{eqnarray*}}
\newtheorem{remark}{Remark}[section]
\title{Computing the ground and first excited states of the fractional Schr\"{o}dinger equation in  an infinite 
potential well\thanks{This research was supported by NSF grant DMS-1217000.}}

\author{ Siwei Duo\thanks{Department of Mathematics and Statistics, Missouri University of Science
and Technology, Rolla, MO
65409-0020, USA ({\tt sddy9@mst.edu}).}\and
Yanzhi Zhang\thanks{Department of Mathematics and Statistics, Missouri University of Science
and Technology, Rolla, MO
65409-0020, USA ({\tt zhangyanz@mst.edu}).}}
\date{}
\begin{document}
\maketitle
\begin{abstract}
In this paper, we numerically study the ground and first 
excited states of the fractional Schr\"{o}dinger equation in an infinite potential well. 
Due to the non-locality of the fractional Laplacian, it is challenging to find the eigenvalues and 
eigenfunctions of the fractional Schr\"{o}dinger equation either analytically or numerically.
We first introduce a fractional gradient flow with discrete normalization and then 
discretize it by using the trapezoidal type quadrature rule in  space and the semi-implicit Euler 
method in time.  Our method can be used to compute the ground and first excited states 
not only in the linear cases but also in the nonlinear cases.
Moreover, it can be generalized to solve 
the fractional partial differential equations (PDEs) with Riesz fractional derivatives in space.
Our numerical results suggest that the eigenfunctions of 
the fractional Schr\"{o}dinger equation in an infinite potential well 
are significantly different from those of the 
standard (non-fractional) Schr\"{o}dinger equation.
In addition, we find that the strong nonlocal interactions represented by the fractional Laplacian can 
 lead to  a large scattering of particles inside of the potential well. Compared to the ground states, 
the scattering of particles in the first excited states is larger.
Furthermore, boundary layers emerge in the ground states and additionally 
inner layers exist in the first excited states  of the fractional nonlinear Schr\"{o}dinger equation. 
 \end{abstract}

%\begin{AMS}
%35Q40, 45C05, 65M06, 81Q05, 82D05
%\end{AMS}

%\pagestyle{myheadings}\thispagestyle{plain}
 %\markboth{S. Duo and Y. Zhang}
%{FRACTIONAL SCHR\"{O}DINGER EQUATION IN BOX POTENTIAL}

% =======================================================================
%                                                          Section 1:  Introduction
% =======================================================================
\section{Introduction}
\setcounter{equation}{0}
\label{section1}

%%%% %%%% %%%% %%%% P1: Fractional Laplacian and fractional Schr\"{o}dinger equation
The fractional Schr\"{o}dinger equation,  a fundamental model of factional quantum mechanics,  
was first introduced by Laskin in \cite{Laskin2000-1, Laskin2000-2, Laskin2002} as the  path integral 
of the L\'{e}vy trajectories.  It is a nonlocal integro-differential equation that is expected to reveal
some novel  phenomena of the quantum mechanics.  Recently,  the fractional Schr\"{o}dinger equation 
in an infinite potential well has attracted massive attention from both physicists and mathematicians,  
and numerous studies have been devoted to finding its eigenvalues and eigenfunctions; see 
\cite{Laskin2000, Banuelos2004, Guo2006, Dong2007, Jeng2010, Kwasnicki2012, Bayin2012, 
Hawkins2012, Bayin2013, Herrmann2013, Uchailkin2013,  Luchko2013, Dong2013} and references 
therein. However, one main debate in the literature is  whether the fractional linear Schr\"{o}dinger 
equation  in an infinite potential well has the same eigenfunctions as those of its standard 
(non-fractional) counterpart,  and so far no agreement has been reached on it \cite{Jeng2010, 
Hawkins2012, Bayin2012, Bayin2013, Dong2013, Luchko2013}. %In addition,  no numerical study 
%can be found  in the literature on directly computing its eigenvalues and eigenfunctions.
The main goal of this paper is to propose an  efficient and accurate numerical method to compute 
the ground and first excited states of   the fractional  Schr\"{o}dinger equation in an infinite 
potential well so as  to advance the understanding on this problem.

%%%% %%%% %%%% %%%% 
We consider the one-dimensional (1D) fractional Schr\"{o}dinger equation of the form 
\cite{Laskin2000, Laskin2002, Jeng2010, Bayin2012, Bayin2013, Kwasnicki2012, 
Hawkins2012, Herrmann2013,  Luchko2013, Kirkpatrick2013}:
\begin{equation}\label{OFNLS}
i\p_t\psi(x,t)={(-\Delta)}^{\ap/2}\psi+V(x)\psi+\beta|\psi|^{2}\psi,\qquad x \in 
{\mathbb R},  \quad \ t > 0, 
\end{equation}
where $\psi(x,t)$ is a complex-valued wave function, and $i = \sqrt{-1}$ denotes the 
imaginary unit.   The constant $\bt\in{\mathbb R}$ describes the strength of the 
local (or short-range) interactions between particles (positive for repulsive interaction 
and negative for attractive interaction), and here we are interested in the case 
of $\bt \geq 0$. $V(x)$ represents the external 
trapping potential, and if an infinite potential well (also known as box potential) is 
considered, it  has the following form: 
\begin{eqnarray}\label{potential}
V(x) = \left\{
\begin{array}{ll}
0, \quad \  & \mbox{if}\ \ |x|<L,\\
\infty,& \mbox{otherwise,}
\end{array}
\right.\qquad x\in{\mathbb R},
\end{eqnarray}
with the constant  $L > 0$.   The Riesz fractional Laplacian $-(-\Dt)^{\ap/2}$ 
is defined through the  principle value integral \cite{Samko1993, Valdinoci0000, 
Stein1970, Du2012, Luchko2013, Delia2013}: 
\bea\label{riesz}
-(-\Dt)^{\ap/2}u(x) &=& -C_{1, \ap} \int_{\mathbb R} \fl{u(x)-u(y)}{|x-y|^{1+\ap}}\,dy \nn\\
&=& C_{1,\ap}\int_{0}^\infty  \fl{u(x-\xi) - 2u(x) + u(x+\xi)}{\xi^{1+\ap}}\, d\xi, \qquad 
x\in{\mathbb R}, \quad\ap \in (0, 2),
\eea
where $C_{1,\ap}$ is a normalization constant defined as
\beas
C_{1,\ap} = \fl{2^{\ap-1}\ap\,\Gamma((1+\ap)/2)}{\sqrt{\pi}\,\Gamma(1-\ap/2)}
= \fl{\Gamma(1+\ap) \sin(\ap\pi/2)}{\pi},   \qquad \ap\in(0, 2).
\eeas
It is easy to verify  that the constant $C_{1,\ap} \approx \ap/2$ as $\ap \to 0$, and 
$C_{1,\ap} \approx (2-\ap)$ as $\ap \to 2$ \cite{Vazquez2012, Vazquez2014}.  In fact, the 
fractional Laplacian $-(-\Dt)^{\ap/2}$  can be viewed as a special case of the nonlocal operator 
with the kernel function proportional to ${\mathcal K}(x, y) = 1/|x - y|^{1+\ap}$ 
\cite{Du2012, Delia2013, Tian2013}. It can be also
obtained from the inverse of the Riesz potential  \cite{Landkof1972, Samko1993}.  
%If $\ap = 2$, the 
%equation (\ref{OFNLS}) becomes the standard (non-fractional) Schr\"{o}dinger equation which 
%has been well studied in the literature. For $\ap\in(0, 2)$ but $\ap \neq 1$,  the Riesz fractional Laplacian 
%defined in (\ref{Riesz}) can be also written in terms of of a hypersingular integral \cite{Samko1993, Valdinoci0000, 
%Stein1970, Du2012, Luchko2013, Delia2013}: 
%\bea\label{riesz}
%(-\Dt)^{\ap/2}u(x) = c_{\ap} \int_{\mathbb R} \fl{u(x)-u(y)}{|x-y|^{1+\ap}}\,dy 
%= -c_\ap\int_{0}^\infty  \fl{u(x-\xi) - 2u(x) + u(x+\xi)}{\xi^{1+\ap}}\, d\xi, 
%\eel
%where $c_{\ap} = 1/[2\Gamma(-\ap)\cos(\ap\pi/2)]$ is a universal constant. Unless stated otherwise, 
%we assume that $\ap\in(0, 2)$ but $\ap \neq 1$ for the  fractional Schr\"{o}dinger equation discussed 
%in this paper.

%%%% %%%% %%%% %%%% P2:  Normalization and energy
The fractional Schr\"{o}dinger equation (\ref{OFNLS}) has two conserved quantities:  
the {\it $L^2$ norm, or mass of the wave function}, which we will take to be 
normalized,\begin{equation}\label{norm}
\|\psi(\cdot, t)\|^2 := \int_{\mathbb R}|\psi(x, t)|^2 dx = \int_{\mathbb R}|\psi(x, 0)|^2 dx 
= \|\psi(\cdot,0)\|^2 = 1, \qquad t \geq 0,
\end{equation}
and {\it the total energy}
\bea\label{energy}
 E(\psi(\cdot, t)) &:=& \int_{\mathbb R} \left(\psi^*(-\Dt)^{\ap/2}\psi + V(x)|\psi|^2
+\fl{\bt}{2}|\psi|^4\right) dx\nn\\
&=& E(\psi(\cdot, 0)), \qquad t\geq0,
\eea
where $f^*$ represents  the complex conjugate of a function $f$. 

%%%% %%%% %%%% %%%% 
In some literature of fractional partial differential equations (PDEs), the  Riesz fractional 
Laplacian $-(-\Dt)^{\ap/2}$  is also defined via a pseudo-differential operator with the symbol 
$-|\kappa|^\ap$  \cite{Samko1993, Laskin2000, Zoia2007, Yang2010}:
\bea\label{Riesz}
-(-\Dt)^{\ap/2} u(x) = {\mathcal F}^{-1}(-|\kappa|^\ap{\mathcal F}(u)), \qquad x\in{\mathbb R},
\eea
where  
\beas
{\mathcal F}(u)(\kappa) = \int_{{\mathbb R}} u(x)\, e^{-i\kappa x} dx, \qquad \kappa \in {\mathbb R}
\eeas
defines the Fourier transform of a function $u(x)$, and ${\mathcal F}^{-1}$ represents the inverse 
Fourier transform.  We remark that {\it if the function $u(x)$ belongs to the Schwartz space,
  the integral representation of the Riesz fractional Laplacian 
$-(-\Dt)^{\ap/2}$ in (\ref{riesz}) is equivalent to its pseudo-differential representation in (\ref{Riesz}) 
\cite{Samko1993, Vazquez2012, Stein1970, Valdinoci0000}.} 
However, if  $u(x)$ is not defined in the Schwartz space,  no report can be found yet in the literature
on the equivalence of (\ref{riesz})  and (\ref{Riesz}),  and it is beyond the scope of this paper.

%%%% %%%% %%%% %%%% P3: Literature review
In this paper, we are interested in computing the eigenvalues and eigenfunctions of the fractional 
Schr\"{o}dinger equation (\ref{OFNLS})--(\ref{potential}) with the fractional Laplacian defined in (\ref{riesz}). 
In physics literature, the  eigenfunctions are usually called the {\it stationary states}. 
In particular, the eigenfunction with the smallest nonzero eigenvalue is called the 
{\it ground state},  and those with the larger eigenvalues are called the  {\it excited states}. 
%So far, most studies on the eigenvalues and eigenfunctions focus on the linear case with 
%$\bt = 0$. If $\ap = 2$,  
In the linear (i.e., $\bt = 0$) case, the eigenvalues and eigenfunctions of the standard
(non-fractional)  Schr\"{o}dinger equation 
in an infinite potential well can be found exactly; see \cite{Greiner, Bao2007, Zhang} and our review in Section \ref{section2-1}. 
However,  the eigenvalues and eigenfunctions of the fractional Schr\"{o}dinger equation  
in an infinite potential well are not well understood yet, and its discussion  in the literature
can be mainly classified into two categories based on the representation of the Riesz fractional 
Laplacian $-(-\Dt)^{\ap/2}$.

%%%% %%%% %%%% %%%% 
On the one hand,  numerous arguments have been made based on the pseudo-differential 
representation of the fractional Laplacian $-(-\Dt)^{\ap/2}$  in (\ref{Riesz}) \cite{Laskin2000, 
Laskin2000-1, Laskin2000-2,  Guo2006, Dong2007, Wang2007, Jeng2010,  Bayin2012, 
Bayin2013}. For instance,  Laskin first stated in \cite{Laskin2000, Laskin2000-1, Laskin2000-2} 
that the eigenfunctions of the fractional linear Schr\"{o}dinger equation in an infinite potential well 
are identical to those of the standard Schr\"{o}dinger equation, while the eigenvalues are modified with 
a power $\ap$. Since then, these results have been used  in many studies of  the fractional partial 
differential equations \cite{Dong2007, Wang2007, Yang2010, Bayin2012, Bayin2013}. 
However,  Jeng et al. recently pointed out in \cite{Jeng2010} that the method used by Laskin 
in finding the eigenfunctions of the fractional Schr\"{o}dinger equation is invalid,  as it 
fails to take the nonlocal character of the fractional Laplacian into account.  They further proved by 
contradiction that the eigenfunctions of  the standard Schr\"{o}dinger equation in an infinite potential 
well can not serve as the eigenfunctions in the fractional cases, and hence the eigenvalues and 
eigenfunctions of the fractional Schr\"{o}dinger equation obtained in \cite{Laskin2000, 
Laskin2000-1, Laskin2000-2, Guo2006, Dong2007} are incorrect. Later,  Bayin argued in
 \cite{Bayin2012}   that the proof presented by Jeng et al. in \cite{Jeng2010} 
was not true, implying that Laskin's solutions are correct.   While in  \cite{Hawkins2012} Hawkins and 
Schwartz criticized  Bayin's calculation and provided a correction to it. 
So far, the controversy over the eigenvalues and eigenfunctions of the fractional linear
Schr\"{o}dinger equation in an infinite potential well is still continuing \cite{Kwasnicki2012, 
Bayin2013, Uchailkin2013, Herrmann2013, Luchko2013, Dong2013}.
%%%% Definition of PV integral
On the other hand, some studies have been reported based on  the integral representation of the 
fractional Laplacian $-(-\Dt)^{\ap/2}$ in (\ref{riesz}).  For example,  Luchko reformulated the fractional 
Schr\"{o}dinger equation in terms of three integral equations and concluded that the results 
by Laskin and many other authors cannot be valid \cite{Luchko2013}.  Herrmann made the same 
conclusion in \cite{Herrmann2013}. 

%%%% %%%% %%%% %%%% %%%% 
%To the best of our knowledge,   the eigenvalues and eigenfunctions 
%of the fractional  Schr\"{o}dinger equation in an infinite potential well still remain an open question, 
Due to the non-locality of the fractional Laplacian, it is very challenging to obtain the analytical 
solutions to the eigenvalues and eigenfunctions of the fractional Schr\"{o}dinger equation in an 
infinite potential well. 
In \cite{Banuelos2004}, Ba$\tilde{\rm n}$uelos provided an estimate on the lower and upper 
bounds for the smallest eigenvalue of the linear Schr\"{o}dinger equation for $\ap \in (0, 2]$. 
Later, a more general estimate was obtained in \cite{Deblassie2004, Chen2005} for all eigenvalues.
In \cite{Kwasnicki2012}, Kwa\'{s}nicki presented the asymptotic approximations to all eigenvalues of 
the fractional linear Schr\"{o}dinger equation in a bounded domain. 
Compared to the eigenvalues, the results on the eigenfunctions of the fractional Schr\"{o}dinger 
equation are very limited.   In \cite{Zoia2007},  Zoia et al. studied  the eigenfunctions 
by using the discrete fractional Laplacian. Luchko conjectured in 
\cite{Luchko2013} that the eigenfunctions of the fractional  Schr\"{o}dinger equation in an 
infinite potential well cannot be expressed in terms of the  elementary  functions. 
Surprisingly,  no numerical studies have been reported on the eigenvalues and eigenfunctions by 
directly solving the fractional Schr\"{o}dinger equation. 
In fact, one main challenge in numerically solving the eigenvalues and eigenfunctions 
is the discretization of the fractional Laplacian $-(-\Dt)^{\ap/2}$. 

%%%% %%%% %%%% %%%% %%%%
In this paper, we propose an  efficient and accurate numerical method to compute  the ground and first excited 
states of the fractional Schr\"{o}dinger equation in an infinite potential well and  attempt to provide 
some insights into their analytical solutions. We remark that  {\it our numerical method and results 
reported in this paper are based on the integral representation of the Riesz fractional Laplacian 
$-(-\Dt)^{\ap/2}$ in (\ref{riesz}) for $\ap\in(0, 2)$}. 
%We will leave its relation 
%to the eigenvalues and eigenfunctions with the pseudo-differential reentation of $-(-\Dt)^{\ap/2}$   for the future work.
Our main contributions in this paper can be briefly summarized as follows. 

\begin{enumerate}
%\item[(i)] We compare the Riesz fractional Laplacian 
%$-(-\Dt)^{\ap/2}$ defined in (\ref{riesz}) with the standard non-fractional Laplacian $\Dt$ 
%and present one example to show that the eigenfunctions of the standard Schr\"{o}dinger equation 
%in an infinite potential well cannot serve as the eigenfunctions in the fractional cases.  Our results 
%are consistent with the conclusions made in \cite{Luchko2013, Herrmann2013}.
%We conclude that the eigenvalues and eigenfunctions found in
%\cite{Laskin2000, Laskin2000-1, Laskin2000-2, Dong2007, Bayin2012, Bayin2013, Dong2013}
%are incorrect.

\item[(i)] To compute the ground and first excited states, we introduce a fractional gradient flow with 
discrete normalization and then propose a novel numerical method to solve it -- a method using the 
trapezoidal type quadrature rule method in space and the semi-implicit Euler method in time. 
Our method can be used to find the ground and first excited states for both linear and nonlinear 
Schr\"{o}dinger equations. Furthermore, it can be generalized to solve the fractional partial 
differential equations (PDEs) with Riesz fractional derivatives in space \cite{Duo2014}.

\item[(ii)] We numerically find  the ground and first excited states and their corresponding eigenvalues 
of the fractional linear ($\bt = 0$) Schr\"{o}dinger equation. The nonlocal interactions from the 
fractional Laplacian lead to a large scattering of particles inside of the potential well, and the smaller 
the parameter $\ap$, the larger the scattering.  In addition, 
%we compare our numerical results with 
%the analytical estimates of the eigenvalues in \cite{Banuelos2004, Deblassie2004, Chen2005, 
%Kwasnicki2012}, and 
our simulated eigenvalues are consistent with the lower and upper bound 
estimates provided in \cite{Banuelos2004, Deblassie2004, Chen2005} as well as the asymptotic 
approximations obtained in \cite{Kwasnicki2012}, showing that our method is accurate in computing 
the ground and first excited states.

\item[(iii)] We study the ground and first excited states of the fractional nonlinear 
Schr\"{o}dinger equation with $\bt > 0$. 
%The competition of the local and nonlocal interactions 
%on the ground and first  excited states are investigated.  
We  find that the presence of the local nonlinear interactions lead to boundary layers in the ground 
states, and additionally inner layers in the excited states. The width of the 
boundary layers depends on both the parameters $\ap$ and $\bt$. 
\end{enumerate}

%%%% %%%% %%%% %%%% P4: Organization
The rest of this paper is organized as follows.  In Section \ref{section2},  we first review some 
analytical results on the eigenvalues and eigenfunctions of the standard Schr\"{o}dinger equation, 
and then reformulate the eigenvalue problem  of the fractional Schr\"{o}dinger equation by taking the
 non-locality of $-(-\Dt)^{\ap/2}$ into account. 
% One example is presented to show the differences 
%between the standard and the Riesz fractional  Laplacian.  
In Section \ref{section3}, we propose a numerical method 
to compute the ground  and  first excited states of fractional Schr\"{o}dinger equation. 
The discretization of the Riesz fractional Laplacian is described in detail.  The ground states and the first excited 
states are reported and discussed in Sections \ref{section4} and \ref{section5} in the linear $(\bt = 0)$ 
and nonlinear ($\bt \neq 0$) cases, respectively. In Section \ref{section6}, we make some conclusions and 
discussions.

% =======================================================================
%                                                          Section 2:  
% =======================================================================
\section{Stationary states in an infinite potential well}
\setcounter{equation}{0}
\label{section2}

%%%% %%%% %%%% %%%% P1: Stationary states
To find the stationary states of  (\ref{OFNLS}), we write the wave function in the form:
\bea\label{ansatz}
\psi(x, t) = e^{-i\mu t}\phi(x),\qquad x\in{\mathbb R},\quad  \  t\geq 0,
\eea
where $\mu \in {\mathbb R}$. 
Substituting the ansatz (\ref{ansatz}) into (\ref{OFNLS}) and taking the mass conservation 
(\ref{norm}) into account, we obtain the following time-independent fractional 
Schr\"{o}dinger equation:
\bea\label{eig}
&&\mu\,\phi(x)  = (-\Dt)^{\ap/2}\phi + V(x)\phi + \bt|\phi|^2\phi, \qquad  x\in{\mathbb R}
\qquad\qquad
\eea
with the constraint
\bea
\label{constraint}
&&\|\phi\|^2 = \int_{\mathbb R} |\phi(x)|^2 dx = 1.
\eea
This is a constrained eigenvalue problem, and the 
eigenvalue $\mu$ (also called {\it chemical potential}) can be calculated from its 
corresponding eigenfunction $\phi(x)$ via:
\bea\label{mu}
\mu = \mu(\phi) &:=& \int_{\mathbb R} \left(\phi^*(-\Dt)^{\ap/2}\phi + V(x)|\phi|^2
+\bt|\phi|^4\right) dx \nn\\
&=& E(\phi) + \fl{\bt}{2} \int_{\mathbb R} |\phi|^4 dx.\qquad
\eea
In fact, the eigenfunctions of (\ref{eig}) under the constraint (\ref{constraint}) are equivalent to 
the critical points of the energy $E(\phi)$ over the set ${\mathcal T} = \{\phi(x) \,|\, 
\mbox{$\|\phi\|^2 = 1$ and $E(\phi) < \infty$} \}$. 

%%%% %%%% %%%% %%% 
Let $\Og = (-L, L)$ denote the bounded domain where the potential $V(x) \equiv 0$. 
For $x\in{\mathbb R}\backslash\Og$, the potential $V(x) = \infty$, and consequently the wave 
function  $\phi(x)\equiv 0$ for any $x\in{\mathbb R}\backslash\Og$,  
since the mass $\|\phi\|^2 = 1$ and  the energy  $E(\phi) < \infty$.  Hence, the problem 
solving for the eigenfunctions of the Schr\"{o}dinger equation in an infinite potential well 
is reduced to finding the eigenfunction $\phi(x)$ in the bounded domain $\Og$ together with
$\phi(x)\equiv 0$ for $x\in{\mathbb R}\backslash\Og$. The corresponding eigenvalue 
can be calculated by
\bea
\mu = \mu(\phi) &:=& \int_{\mathbb R} \left(\phi^*(-\Dt)^{\ap/2}\phi
+\bt|\phi|^4\right) dx.\qquad
\eea
In the following,  we will focus  on finding the eigenfunctions of  the eigenvalue problem 
(\ref{eig})--(\ref{constraint}) for $x\in\Og$.
 
% -----------------------------------------------------------------------------------------------------------------------------
%                                                              Section 2-1: SNLSE
% -----------------------------------------------------------------------------------------------------------------------------
\subsection{Standard Schr\"odinger equation}
\label{section2-1}

%%%% %%%% %%%% %%%% P1: 
For the convenience of readers, we briefly review the eigenvalues and eigenfunctions of 
the standard Schr\"{o}dinger equation in this section. First, we present their exact solutions 
in the linear $(\bt = 0)$ cases.  In the nonlinear cases with $\bt \gg 1$,  we obtain  
the leading-order approximations to  the eigenvalues and eigenfunctions.  These analytical results 
can be used to compare with those of the fractional Schr\"{o}dinger equation so as 
to understand the differences between the standard and fractional Schr\"{o}dinger equation.

%%%% %%%% %%%% %%%% P2: 
%If $\ap = 2$, the Laplacian $(-\Dt)^{\ap/2} = -\Dt$, 
Replacing the fractional Laplacian $-(-\Dt)^{\ap/2}$ with the standard Laplacian  $\Dt$,  
the eigenvalue problem (\ref{eig}) reduces to the  standard 
Schr\"{o}dinger equation. The eigenfunction $\phi(x)$ for $x\in\Og$ can be found by solving the bounded 
problem \cite{Greiner, Jeng2010, Bao2007}:
\bea
\label{eig1}
&&\mu \phi(x)  = -\Dt\phi(x) + \bt|\phi(x)|^2\phi(x), \qquad  x\in\Og,
\eea
along with the homogeneous Dirichlet boundary conditions
\bea\label{FBC1}
\phi(L) = \phi(-L) =  0,
\eea
and the constraint of normalization
\bea
\label{norm1}
\|\phi\|^2 = \int_{-L}^{L}|\phi(x)|^2 dx = 1.
\eea
Note that the two-point homogeneous Dirichlet boundary conditions are applied in 
(\ref{eig1})--\eqref{norm1},   due to the fact that the standard Laplacian $\Dt$ is a local operator 
and the wave function $\phi(x)$ is continuous at $x = \pm L$. That is,  the eigenvalues and 
eigenfunctions of the standard 
Schr\"{o}dinger equation can be solved in a piecewise approach -- finding the solutions inside of 
the infinite potential well  and then using their continuity to match up with those outside  the potential 
well.

%%%% %%%% %%%% %%%% P3:
In the linear (i.e., $\bt = 0$) cases,  the eigenvalues and eigenfunctions of  (\ref{eig1})--(\ref{norm1}) 
can be found exactly. For $x\in\Og$,  the $s$-th eigenfunction has the form 
 \cite{Greiner, Laskin2000, Bao2007}:
\bea\label{phis0}
\phi_s(x) = \sqrt{\fl{1}{L}}\sin\left[\fl{(s+1)\pi}{2}\Big(1+\fl{x}{L}\Big)\right], \quad \  x\in\Og, \qquad 
s\in{\mathbb N}^0,
\eea
and the corresponding eigenvalue is
\bea\label{mu0}
\mu_s := \mu(\phi_s) = \left[\fl{(s+1)\pi}{2L}\right]^2, \qquad s\in{\mathbb N}^0,
\eea 
where the ground states and the first excited states correspond to $s = 0$ and $s = 1$, respectively. 

%%%% %%%% %%%% %%%% P3:
In the nonlinear cases with $\bt\neq 0$,  the constrained eigenvalue problem  (\ref{eig1})--(\ref{norm1}) 
cannot be exactly solved.  However,  the results in (\ref{phis0})--(\ref{mu0}) provide 
a good approximation to the eigenfunctions and eigenvalues in the weakly interacting regimes with 
$\bt\sim o(1)$. In 
the strongly repulsive interacting cases  (i.e., $\bt \gg 1$),  we can 
 find the leading-order approximation 
(also called {\it Thomas--Fermi approximation})  of the $s$-th ($s\in{\mathbb N}^0$) eigenfunction 
\cite{Zhang, Bao2007}:
\bea\label{eigen2}
&&\phi_s(x) \approx \phi_s^a(x) = \sqrt{\fl{\mu_s^a}{\bt}}\Bigg\{\sum_{r = 0}^{\lfloor(s+1)/2\rfloor} 
\tanh\bigg[\fl{\sqrt{2\mu_s^a} L}{2}\bigg(\Big(1+\fl{x}{L}\Big)-\fl{4r}{s+1}\bigg)\bigg]\qquad\nn\\
&&\qquad \quad +\sum_{r=0}^{\lfloor s/2\rfloor}\tanh\bigg[\fl{\sqrt{2\mu_s^a} L}{2}\bigg(\fl{4r+2}{s+1}
-\Big(1+\fl{x}{L}\Big)\bigg)\bigg] - c_s\tanh\left(\fl{\sqrt{2\mu_s^a}L}{2}\right)\Bigg\},\quad\ x\in\Og, 
\eea
where $\lfloor r \rfloor$ defines the floor function of a real number $r$, and the constant 
\beas
c_s = \left\{\begin{array}{ll}
1, \quad &\mbox{if $s$ is even,}\\
0, &\mbox{if $s$ is odd}.
\end{array}\right.
\eeas
Correspondingly, the leading-order approximation of the eigenvalue is
\bea
\mu_s\, \approx\, \mu_s^a = \fl{1}{L^2}\left[\fl{L}{2}\bt + (s+2)\sqrt{\bt L+ (s+2)^2} + (s+2)^2\right], 
\qquad s\in{\mathbb N}^0.
\eea
The approximations in (\ref{eigen2}) show that when $\bt \gg 1$, all the stationary states
of the standard nonlinear Schr\"{o}dinger equation have boundary layers. In addition, for 
$s \geq 1$,   the excited states also have inner layers,  and the number of inner layers in 
the $s$-th excited state is equal to $s$.

% -----------------------------------------------------------------------------------------------------------------------------
%                                                              Section 2-2: FNLSE
% -----------------------------------------------------------------------------------------------------------------------------
\subsection{Fractional Schr\"{o}dinger equation}
\label{section2-2}

%%%% %%%% %%%% %%%% P1:
In contrast to the standard Schr\"{o}dinger equation, the stationary states of the fractional 
Schr\"{o}dinger equation in an infinite potential well have not been well understood yet. 
%In \cite{Laskin2000},  the stationary states of the fractional 
%Schr\"{o}dinger equation were first discussed based on the pseudo-differential representation 
%of $-(-\Dt)^{\ap/2}$, and the eigenvalues and eigenfunctions were solved in the same manner as used in
%the standard cases -- solving the bounded eigenvalue problem (\ref{eig1})--(\ref{norm1}) but replacing the 
%operator $-\Dt$ with $(-\Dt)^{\ap/2}$. It was claimed in \cite{Laskin2000} 
%that the fractional Schr\"{o}dinger equation in an infinite potential well has the same eigenfunctions 
%as those of its standard counterpart, but the eigenvalues are modified with a power 
%$\ap$, i.e., $\mu_s  = [(s+1)\pi/2L]^\ap$ in fractional cases. 
%Since then,  these results have been used as the benchmark problem in many studies; see, e.g.,
 %\cite{Laskin2000-1, Laskin2000-2,  Laskin2002,  Wang2007, Yang2010, Bayin2012, Dong2013} 
 %and references therein. 
%Recently,  in \cite{Jeng2010} Jeng et al. pointed out  that the eigenfunctions of the standard 
%Schr\"{o}dinger equation in an infinite well potential can not be the eigenfunctions in the 
%fractional  cases, and they presented a proof by contradiction.  Later, Luchko in \cite{Luchko2013} 
%and Herrmann in \cite{Herrmann2013} made the same conclusions by using  different approaches. 
%However, they did not provide the correct solutions,  and the eigenvalues and eigenfunctions of the 
%fractional Schr\"{o}dinger equation in an infinite potential well are still an open question \cite{Jeng2010, 
%Kwasnicki2012, Luchko2013, Herrmann2013, Uchailkin2013}.
%%%% %%%% %%%% %%%% P2:
Our main goal of this work is to compute the ground and first excited states of the fractional 
Schr\"{o}dingier equation in an infinite potential well so as to advance the understanding on 
this problem.  Unlike the standard Laplacian, the fractional Laplacian describing the long-range 
interactions is a  nonlocal operator, that is,  the function $(-\Dt)^{\ap/2}\phi(x)$ depends  on the 
wave function $\phi(y)$ not only for $y\in\Og$ but also for $y\notin \Og$, albeit $\phi(y) \equiv 0$ 
when $y\notin\Og$. As a result, in the fractional cases,  the eigenvalue problem 
\eqref{eig}--\eqref{constraint} can not be truncated into a bounded domain. In fact,  it is straightforward 
to consider the following eigenvalue problem:
\bea
\label{eig2}
&&\mu \phi(x)  = (-\Dt)^{\ap/2}\phi(x) + \bt|\phi(x)|^2\phi(x), \qquad  x\in\Og,
\eea
with the {\it nonzero volume constraint} \cite{Du2012, Tian2013, Guan0000}
\bea\label{FBC2}
\phi(x) = 0, \qquad
x\in{\mathbb R}\backslash\Og, 
\eea
and the normalization constraint (\ref{norm1}). 
%\bea
%\label{norm2}
%\|\phi\|^2 = \int_{-L}^L |\phi(x)|^2\,dx = 1.
%\eea
Here,  we take the nonlocal character of the fractional Laplacian into account 
and apply the nonlocal boundary condition (\ref{FBC2}) to the time-independent Schr\"{o}dinger 
equation (\ref{eig2}).  %{\color{red}In the standard non-fractional cases,  $\Dt$ is a local operator,  and 
Due to the non-locality, it is very challenging to solve (\ref{eig2})--(\ref{FBC2}) analytically, 
 and thus the analytical solutions to eigenvalues and eigenfunctions 
 still remain an open question.  So far, only some estimates 
and asymptotic approximations of the eigenvalues are  reported in the literature for the linear  
($\bt =0$)  cases \cite{Banuelos2004,  Deblassie2004, Chen2005, Kwasnicki2012}.  For the 
convenience of readers,  we review the main results in the following remarks: 

%%%% %%%% %%%% %%%% Remark 1
\begin{remark} {\bf (Lower and upper bounds of  the eigenvalues) } 
Various estimates can be found  in \cite{Banuelos2004, Deblassie2004, Chen2005} for 
the lower and upper bounds of  the eigenvalue $\mu_s$ of 
the fractional linear $(\bt = 0)$ Schr\"{o}dinger equation in an interval of length $l$. 
For $s\in{\mathbb N}^0$,  the lower and upper bounds of the eigenvalue $\mu_s$ 
are (\cite[p. 9]{Chen2005}): 
\bea\label{eq00}
\fl{1}{2}\left[\fl{(s+1)\pi}{l}\right]^\ap \leq \mu_s \leq \left[\fl{(s+1)\pi}{l}\right]^\ap, \qquad 
\ap\in(0, 2],\quad \ \ s\in{\mathbb N}^0
\eea
While in \cite{Banuelos2004},  a different estimate was provided for the 
smallest eigenvalue (i.e., $s = 0$, corresponding to  ground state)  (
\cite[Corollary 2.2]{Banuelos2004}):
\bea\label{eq4}
\qquad p(\ap)\, \leq\, \mu_0\, \leq\, p(\ap)\, \fl{{B}(\fl{1}{2}, 1+\fl{\ap}{2})}{{B}(\fl{1}{2}, 1+\ap)}\ 
\quad\mbox{with}\quad p(\ap) = \fl{2^\ap\,\Gamma(1+\fl{\ap}{2})\,\Gamma(\fl{1+\ap}{2})}
{\Gamma(\fl{1}{2})},
\eea
for $\ap\in(0, 2]$, where $B(a, b)$ defines the Beta function of  $a$ and $b$. 
It is easy to verify that when $s = 0$,  the estimates in (\ref{eq4}) are much sharper than 
those in (\ref{eq00}), but the estimate in (\ref{eq00}) is valid for any $s\in{\mathbb N}^0$.
\end{remark}

%%%% %%%% %%%% %%%% Remark 2
\begin{remark} {\bf (Asymptotic approximations of the eigenvalues)} When $\bt = 0$,  the asymptotic 
approximation of the $s$-th eigenvalue  of the fractional linear Schr\"{o}dinger equation in an 
interval $(-1, 1)$ is (\cite[Theorem 1]{Kwasnicki2012}): 
\bea\label{eq5}
\qquad \mu_s = \left[\fl{(s+1)\pi}{2} - \fl{(2-\ap)\pi}{8}\right]^\ap + O\left(\fl{2-\ap}{(s+1)\sqrt{\ap}}\right),
 \quad \ \ap\in(0, 2],  
\quad s \in {\mathbb N}^0.
\eea
In fact, when $\ap = 2$, (\ref{eq5}) gives the  exact eigenvalue $\mu_s = [(s+1)\pi/2]^2$ (for
$s\in{\mathbb N}^0$) of the standard linear Schr\"{o}dinger equation in an infinite potential well.
\end{remark}

%%%% %%%% %%%% %%%% P
For the smallest eigenvalue $\mu_0$, it is easy to verify  when $1\leq \ap \leq 2$ 
the estimates obtained in \cite{Banuelos2004} are 
consistent with the asymptotic approximations provided in \cite{Kwasnicki2012}.  However,  when $0<\ap < 1$,  the asymptotic results in \cite{Kwasnicki2012} are 
always larger than the upper bound presented in \cite{Banuelos2004}. 
In Section  \ref{section4}, we will compare these results with our numerical results (see Table \ref{T1}) and provide more discussions.

%%%% %%%% %%%% %%%% P
Even though the estimates of eigenvalues are obtained in \cite{Banuelos2004, Deblassie2004, 
Chen2005, Kwasnicki2012}, the information on the eigenfunctions is still very limited.  
In \cite{Luchko2013}, Luchko conjectured  that 
the eigenfunctions of the  fractional Schr\"{o}dinger equation cannot be written in terms of elementary 
functions.  In \cite{Zoia2007},  Zoia et al. studied the eigenfunctions of the discrete fractional Laplacian. 
The existence and uniqueness of the ground states solution of the general fractional Schr\"{o}dinger equation 
can be found in \cite{Feng2013, Chang2013, Secchi2013}. 
Surprisingly,  no study has been carried out  by directly 
simulating the  fractional Schr\"{o}dinger equation, and furthermore no results can be found in the literature 
on the stationary states of the fractional nonlinear ($\bt\neq 0$) Schr\"{o}dinger equation in an infinite 
potential well.

% =======================================================================
%                                                          Section 3: 
% =======================================================================
\section{Fractional gradient flow and its discretization}
\setcounter{equation}{0}
\label{section3}

%%%% %%%% %%%% %%%% P1
In this section, we propose an efficient and accurate method for computing the ground and first excited 
states of the fractional Schr\"{o}dinger equation in an infinite potential 
well. First, we introduce the fractional gradient flow with  discrete normalization (FGFDN), 
analogous to the normalized  gradient flow used in finding the stationary states of the standard 
Schr\"{o}dinger equation \cite{Bao2004, Chiofalo2000}.  
Then we discretize the FGFDN by using the trapezoidal type quadrature rule method in  space  
and the semi-implicit Euler method in time.
Our method  can be used to find the ground and first excited states of both linear 
and nonlinear fractional Schr\"{o}dinger equation in an infinite potential well.
Furthermore,  it can be generalized to study the partial differential equations (PDEs) with 
Riesz fractional derivatives in space. 

%%%% %%%% %%%% %%%% P2: 
Let $\Dt t> 0$ denote the time step, and then define the time sequence  $t_n = n\Dt t$ for $n = 0, 1, \ldots$.  
From time $t = t_n$ to $t = t_{n+1}$,  the fractional gradient flow with  discrete normalization 
(FGFDN)  is given by:
\begin{eqnarray}\label{GFDN}
&&\qquad\ \fl{\p\phi(x,t)}{\p t}=C_{1,\ap}\int_0^\infty\fl{\phi(x-\xi, t) -2\phi(x,t) + \phi(x+\xi,t)}{\xi^{1+\ap}}\,d\xi 
- \bt|\phi(x, t)|^2\phi(x, t), \nn\\%\quad x\in\Og,\qquad\quad\\
&&\qquad\qquad\qquad\qquad\qquad\qquad\qquad\qquad\qquad\qquad\qquad\qquad
\quad  \  \ x\in\Og,\quad t_n\leq t\leq  t_{n+1},\\
\label{eq000}
&&\qquad \  \phi(x, t) = 0, \qquad x\in{\mathbb R}\backslash\Og,\quad t_n\leq t\leq  t_{n+1},
\eea
and at the end of each time step, the wave function $\phi(x,t)$ is projected to satisfy the 
normalization condition in (\ref{norm1}):
\bea
\label{GFDN-BC}
\phi(x,t_{n+1}) =\frac{\phi(x,t_{n+1}^{-})}{\|\phi(\cdot, t_{n+1}^{-})\|},\qquad x\in\Og \quad 
\eea
with $\phi(x, t_{n+1}^-)$ the solution obtained from (\ref{GFDN})--(\ref{eq000}) at $t = t_{n+1}$,  and 
the norm  $\|\,\cdot\, \| = \|\,\cdot\,\|_{l^{2}(\Og)}$. The initial condition at time $t = 0$ is given by
\bea\label{GFDN-Initial}
&&\phi(x, 0) = \varphi(x), \quad x\in\Og, \quad\mbox{with}\quad \|\varphi\| = 1,  \qquad \\
 &&\phi(x, 0) = 0, \quad x\in{\mathbb R}\backslash\Og.
\eea
In fact, the FGFDN (\ref{GFDN})--(\ref{GFDN-BC})  can be viewed as first applying the steepest 
decent method to the energy functional (\ref{energy}) and then projecting the solution 
back to satisfy the normalization constraint (\ref{norm}). For more discussions 
on normalized gradient flow, see \cite{Bao2004} and references therein. 

%%%% %%% %%%% %%%% 
As  previously discussed, the Riesz fractional Laplacian is a nonlocal operator defined 
in the whole space ${\mathbb R}$, that is, at any point $x\in{\mathbb R}$,  the wave  function $\phi(x,t)$ 
interacts with  $\phi(y,t)$ for all $y\in{\mathbb R}$ but $y \neq x$. 
However,  the strength of their interactions is proportional to $1/|x-y|^{1+\ap}$, decaying
 as the distance $|x - y|$ increases.  Choosing a constant $A \geq 2L$,  we can rewrite the 
 integral in \eqref{GFDN} as
\bea\label{opL}
{\mathcal L}_0^\infty \phi(x,t)&:=&C_{1,\ap}\int_0^\infty\fl{\phi(x-\xi, t) -2\phi(x,t) + \phi(x+\xi,t)}{\xi^{1+\ap}}\,d\xi 
\nn\\ &=& {\mathcal L}_0^A\phi(x,t)  + {\mathcal L}_A^\infty \phi(x,t),
\eea
where the operator ${\mathcal L}_a^b$ is defined by
\beas
{\mathcal L}_a^b \phi(x,t) := C_{1,\ap}\int_a^b\fl{\phi(x-\xi, t) -2\phi(x,t) + \phi(x+\xi,t)}{\xi^{1+\ap}}\,d\xi.
\eeas
Since $A \geq 2L$, we obtain $(x\pm \xi)\notin\Og$ for any points $x\in\Og$ and $\xi \geq A$, 
and consequently  the wave function $\phi(x\pm \xi, t) \equiv 0$. Hence,  the integral 
${\mathcal L}_A^\infty\phi(x,t)$ reduces to:
 \bea
{\mathcal L}_A^\infty\phi(x,t) &:=& C_{1,\ap}\int_A^\infty\fl{\phi(x-\xi, t) -2\phi(x, t) 
+ \phi(x+\xi, t)}{\xi^{1+\ap}}\,d\xi\nn\\
\label{eq66}
&=& -2\,C_{1,\ap} \phi(x,t)\int^\infty_A\fl{1}{\xi^{1+\ap}} \, d\xi = -2\fl{C_{1,\ap}}{\ap A^\ap}\phi(x,t), 
\qquad x\in\Og,\qquad\quad
\eea
that is,  ${\mathcal L}_A^\infty$ can be integrated exactly if choosing $A\geq 2L$.

%%%% %%%% %%%% %%%%
Next, we focus on evaluating the integral ${\mathcal L}_0^A\phi(x,t)$ numerically. First, we write it 
in the following form:
\bea
{\mathcal L}_0^A\phi(x,t) &:=& C_{1,\ap}\int_0^A\fl{\phi(x-\xi, t) -2\phi(x, t) + \phi(x+\xi, t)}{\xi^{1+\ap}}d\xi\nn\\
\label{eq6}
&=& C_{1,\ap}\int_0^A\fl{\phi(x-\xi,t) -2\phi(x,t) + \phi(x+\xi,t)}{\xi^{2-\gm}}\cdot \fl{1}{\xi^{-1+(\ap+\gm)}}\, d\xi, 
\qquad x\in\Og,\qquad \nn
\eea
where the constant $\gm\in(0, 2-\ap)$, and the selection of $\gm$ will be discussed in Remark 
\ref{remark3-2}. 
%In \eqref{opL}, we divide the integral  into two parts: ${\mathcal L}_0^A\phi(x,t)$ and 
%${\mathcal L}_A^\infty\phi(x,t)$, and the latter can be found exactly. Hence, 
%and in the following, we will focus on the numerical approximation of ${\mathcal L}_0^A\phi(x,t)$.
%%%% %%%% %%%% %%%% P1
Without loss of generality, we choose the constant $A = 2mL$ for an integer $m\geq1$ and denote 
$\Og_b = \left\{x\,|\, L \leq |x| \leq L+A\right\}$. Let $J$ be a positive
integer. Define the mesh size $h = 2L/J$ and 
the grid points $x_j = -L+jh$ for $j\in{\mathcal S}$,  where the index set 
${\mathcal S}:={\mathcal S}_0 \cup{\mathcal S}_1$ with
${\mathcal S}_0 = \{j\,|\, 1\leq j\leq J-1\}$ and  ${\mathcal S}_1 = \{j\,|\, -M\leq j\leq 0 \ \, \mbox{or} 
\ \, J\leq j\leq M+J \}$.  Here,  the integer $M = mJ$,  i.e.,  ${\mathcal S}_0$ and ${\mathcal S}_1$
 denote the index sets of the grid points in 
$\Og$ and $\Og_b$, respectively. It is easy to verify that $A = Mh$.
At each point $x = x_j$ ($j\in{\mathcal S}_0$),  we can approximate  the 
integral ${\mathcal L}_0^A\phi(x_j,t)$ by:
\bea 
{\mathcal L}_0^{A,h}\phi(x_j,t)&:=&\fl{C_{1,\ap}}{2} \Bigg[\lim_{\xi \to0}\left(\fl{\phi(x_j-\xi, t) - 2 \phi(x_j, t) 
+ \phi(x_j+\xi, t)}{\xi^{2-\gamma}}\right)\int_{0}^{\xi_1} \fl{1}{\xi^{-1+(\ap+\gamma)}}\,d\xi\nn\\
&&\quad+\sum_{l=1}^{M-1}\fl{\phi(x_j-\xi_l,t)-2\phi(x_j, t)+\phi(x_j+\xi_l,t)}{\xi_l^{2-\gm}}
\int_{\xi_{l-1}}^{\xi_{l+1}} \fl{1}{\xi^{-1+(\ap+\gamma)}}\,d\xi\nn\\
\label{eq0}
&&\quad+ \fl{\phi(x_j-\xi_M,t)-2\phi(x_j, t)+\phi(x_j+\xi_M,t)}{\xi_M^{2-\gm}}
\int_{\xi_{M-1}}^{\xi_M} \fl{1}{\xi^{-1+(\ap+\gamma)}}\,d\xi\nn
\Bigg]\nn\\
&=&\fl{C_{1, \ap}}{2} \Big({I}_{0,\,j} + \sum_{l =1}^{M-1}{I}_{l,\,j}  + {I}_{M,\,j}\Big),\qquad j\in{\mathcal S}_0,
\eea
where $\xi_l = lh$ for $0 \leq l \leq M$. Denote $\sigma = 2-(\ap + \gm)$.  We can 
approximate the term ${I}_{0,\,j}$ by:
\bea
{I}_{0,\,j} &=& \lim_{\xi \to0}\left(\fl{\phi(x_j-\xi, t) - 2 \phi(x_j, t) 
+ \phi(x_j+\xi, t)}{\xi^{2-\gm}}\right)\int_{0}^{\xi_1} \fl{1}{\xi^{-1+(\ap+\gamma)}}\,d\xi\nn\\
%&=& \fl{h^\sigma}{\sigma} \lim_{\xi \to0}\left(\fl{\phi(x_j-\xi, t) - 2 \phi(x_j, t) 
%+ \phi(x_j+\xi, t)}{\xi^{2-\gm}}\right)\qquad\qquad\qquad\qquad\nn\\
&=& \fl{h^\sigma}{\sigma} \lim_{\xi \to0}\left(\fl{\phi(x_j-\xi, t) - 2 \phi(x_j, t) 
+ \phi(x_j+\xi, t)}{\xi^{2}}\cdot \xi^{\gamma}\right) \nn\\
&\approx& \fl{h^\sigma}{\sigma}\phi_{xx}(x_j,t) \lim_{\xi\to0} \xi^\gm.
\eea
Assuming that  the wave function $\phi(x,t)$ is smooth enough and $\phi_{xx}(x,t)$ is bounded 
for $x \in \Og$,  we obtain that  ${I}_{0,\, j} \to 0$  \cite{Tian2013}.  In other words,  
the  term ${I}_{0, \, j}$ can be neglected,  if the mesh size $h$ is very small, assuming that 
the function  $\phi_{xx}(x,t)$ is bounded. 
 
 %%%% %%%% %%%% %%%%
Let $\phi_j(t)$ represent the numerical approximation  of $\phi(x_j, t)$.
For $1\leq l\leq M-1$,  we can compute the term ${I}_{l,\,j}$ by:  
\bea
{I}_{l,\,j} &=&  \fl{\phi(x_j-\xi_l, t)-2\phi(x_j, t)+\phi(x_j+\xi_l, t)}
{\xi_l^{2-\gm}}\int_{\xi_{l-1}}^{\xi_{l+1}}\fl{1}{\xi^{-1+(\ap+\gm)}}\,d\xi\nn\\
&=& \fl{\phi(x_j-lh, t)-2\phi(x_j,t)+\phi(x_j+lh, t)}
{(lh)^{2-\gm}}\cdot\fl{\big[(l+1)h\big]^{\sigma}-\big[(l-1)h\big]^{\sigma}}{\sigma}\qquad\qquad\nn\\
&=& \fl{\phi_{j-l}(t)-2\phi_j(t)+\phi_{j+l}(t)}
{l^{2-\gm}}\cdot\fl{(l+1)^{\sigma}-(l-1)^{\sigma}}{\sigma h^\ap},\qquad 1\leq l\leq M-1.\nn
\eea
Note that the nonzero volume constraint in (\ref{eq000}) implies that if $x_{j-l}\in\Og_b$, 
the wave function $\phi(x_{j-l}, t) = 0$, equivalently, we have $\phi_{j-l}(t) = 0$,  if 
$(j-l)\in{\mathcal S}_1$. Hence, 
\bea\label{eqi2}
\sum_{l = 1}^{M-1} I_{l, \, j}&=&\sum_{l=1}^{M-1}\fl{(l+1)^\sigma-(l-1)^\sigma}{\sigma h^\ap\, l^{2-\gm}}
\Big(\phi_{j-l}(t) + \phi_{j+l}(t)\Big)-2\phi_j(t)\sum_{l=1}^{M-1}\fl{(l+1)^\sigma-(l-1)^\sigma}
{\sigma h^\ap\, l^{2-\gm}}\nn\\
&=&\sum_{\substack{k=j-M+1 \\ k\neq j}}^{j+M-1}
\fl{\big(|k-j|+1\big)^\sigma - \big(|k-j|-1\big)^\sigma}{\sigma h^\ap\, |k-j|^{2-\gm}}\phi_k(t)-2\phi_j(t)
\sum_{l=1}^{M-1}\fl{(l+1)^\sigma-(l-1)^\sigma}{\sigma h^\ap\, l^{2-\gm}}\nn\\
&=&\sum_{\substack{k=1 \\ k\neq j}}^{J-1}
\fl{\big(|k-j|+1\big)^\sigma - \big(|k-j|-1\big)^\sigma}{\sigma h^\ap\, |k-j|^{2-\gm}}\phi_k(t)-2\phi_j(t)
\sum_{l=1}^{M-1}\fl{(l+1)^\sigma-(l-1)^\sigma}{\sigma h^\ap \, l^{2-\gm}}. \nn\\
\eea
In (\ref{eqi2}), we have set $k = |j - l|$, equivalently, $\xi_k = |x_j - x_l|$ represents the distance 
between the two points $x_j$ and $x_l$.

%%%% %%%% %%%% %%%% 
Since $A \geq 2L$ and $\xi_M = A$, we have $(x_j \pm \xi_M) \in \Og_b$ for any $x_j\in\Og$. Thus, 
the wave function $\phi_{j-M}(t) = \phi_{j+M}(t) \equiv 0$, and the term ${I}_{M, j}$ can be calculated by:
\bea
{I}_{M, j} &=& \fl{\phi(x_j-\xi_M, t)-2\phi(x_j, t)+\phi(x_j+\xi_M, t)}
{\xi_M^{2-\gm}}\int_{\xi_{M-1}}^{\xi_{M}}\fl{1}{\xi^{-1+(\ap+\gm)}}\,d\xi\nn\\
\label{eqi3}
&=&\fl{\phi_{j-M}(t) - 2\phi_j(t) + \phi_{j+M}(t)}{M^{2-\gm}}\cdot 
\fl{M^\sigma - (M-1)^\sigma}{\sigma h^\ap} \nn\\
&=& -2\,\fl{M^\sigma - (M-1)^\sigma}{\sigma h^\ap M^{2-\gm}}\, \phi_j(t).
\quad\qquad 
\eea
%%%% %%%% %%%% %%%% 
Combining (\ref{opL})--(\ref{eqi3}), we obtain  the numerical 
approximation to the integral ${\mathcal L}_0^\infty\phi(x_j,t)$:
\bea
&&{\mathcal L}_0^{\infty, h}\phi(x_j,t)=\fl{C_{1,\ap}}{2\sigma h^\ap}\sum_{\substack{k=1 \\ k\neq j}}^{J-1}
\fl{\big(|k-j|+1\big)^\sigma - \big(|k-j|-1\big)^\sigma}{|k-j|^{2-\gm}}\phi_k(t)\nn\\
&&\qquad - C_{1, \ap}\phi_j(t)\left(\fl{1}{\sigma h^\ap}\sum_{l=1}^{M-1}
\fl{(l+1)^\sigma-(l-1)^\sigma}{l^{2-\gm}}+\fl{M^\sigma - (M-1)^\sigma}{\sigma h^\ap M^{2-\gm}}
+ \fl{2}{\ap A^\ap}\right), \quad j\in{\mathcal S}_0.\qquad \quad\nn
\eea

%More discussions on numerical discretization of the nonlocal models in space can be found 
%in \cite{Du2012, Delia2013, Tian2013} and references therein.

%%%% %%%% %%%% %%%% 
Let $\Phi(t) = \left(\phi_1(t), \phi_2(t), \ldots, \phi_{J-1}(t)\right)^T$ denote the solution vector 
at time $t$. Then the semi-discretization of the fractional gradient flow in
(\ref{GFDN})--(\ref{eq000}) is given by
\bea\label{semi}
&&\fl{d\Phi(t)}{dt} = {\bf D}\Phi(t) + {\bf F}(\Phi(t)), \, \qquad t\in[t_n, t_{n+1}],
\eea
where the  matrix ${\bf D} = \left\{D_{jk}\right\}_{(J-1)\times(J-1)}$ with 
\bea
{D}_{jk} = -\fl{C_{1,\ap}}{\sigma h^\ap}\left\{\begin{array}{ll}
\displaystyle\sum_{l = 1}^{M-1}\fl{(l+1)^\sigma - (l-1)^\sigma}{l^{2-\gm}}+
\fl{M^\sigma - (M-1)^\sigma}{M^{2-\gm}} + \fl{2\sigma h^\ap}{\ap A^\ap}, \quad  & k = j, \\
\displaystyle \fl{\big(|k-j|+1\big)^\sigma - \big(|k-j|-1\big)^\sigma}
{2|k-j|^{2-\gm}}, \quad \  & k\in{\mathcal S_0} 
\  \mbox{but} \ \,  k\neq j,\\
\end{array}\right. \nn
\eea
for $j\in{\mathcal S}_0$.  We see that   ${\bf D}$ is  a symmetric Toeplitz  matrix.  In addition, it is a full matrix, 
representing the nonlocal characteristic of the fractional Laplacian.
The vector ${\bf F}(\Phi) = \left(f(\phi_1), f(\phi_2), \ldots, 
f(\phi_{J-1})\right)^T$ with the function $f(\phi_j) = -\bt|\phi_j(t)|^2\phi_j(t)$. 

%%%% %%%% %%%% %%%%
The semi-discretization of the fractional gradient flow in (\ref{semi}) is a system of nonlinear 
ordinary differential equations (ODEs).
Denote $\Phi^{n}$ as the numerical approximation of the 
solution vector $\Phi(t_n)$.  We discretize \eqref{semi} in time 
by  the semi-implicit Euler method and obtain the full discretization of the (\ref{GFDN})--(\ref{GFDN-BC}) 
as: 
\bea\label{eq55}
\fl{\Phi^{(1)}- \Phi^{n}}{\Dt t} = {\bf D}\Phi^{(1)} + {\bf F}(\Phi^n), \qquad  n = 0, 1, \ldots.
\eea
and the projection in (\ref{GFDN-BC}) is discretized as 
\bea\label{eq77}
\Phi^{n+1} = \fl{{\Phi}^{(1)}}{\|{\Phi}^{(1)}\|}, \qquad\mbox{with}\quad
\|\Phi^{(1)}\| = \Big(h\sum_{j=1}^{J-1} \big|\phi_j^{(1)}\big|^2\Big)^{1/2}.
\eea
When $n = 0$, the initial condition at $t = 0$ is discretized by
\bea\label{eq88}
\phi_j^0 = \varphi(x_j), \qquad j\in{\mathcal S}_0.
\eea
%%%% %%%% %%%% %%%%
The scheme (\ref{eq55})--(\ref{eq88}) can be used to compute for both the ground states and the 
first excited states of the fractional Schr\"{o}dinger equation in an infinite potential well.
In our simulations, the ground and first excited states are obtained by requiring that
\bea\label{tol}
\fl{\|\Phi^{n+1} - \Phi^n\|_{\infty}}{\Dt t} < \varepsilon
\eea
for a small tolerance $\varepsilon>0$. 

In the following remarks, we will further discuss  the selection 
of the parameters $A$ and $\gm$.
%%%% %%%% %%%% %%%%
\begin{remark} {\bf (Selection of the parameter $A$) } In the scheme, we choose a constant $A$ and 
 rewrite the integral ${\mathcal L}_0^\infty\phi(x,t) = {\mathcal L}_0^A\phi(x,t) 
+ {\mathcal L}_A^\infty\phi(x,t)$.  On the one hand, the selection of  $A$ should 
ensure that the improper integral ${\mathcal L}_A^\infty\phi(x,t)$ 
can be simplified and evaluated by  (\ref{eq66}), which requires that $A \geq 2L$. 
On the other hand,  for a fixed mesh size $h$,  we want to have 
the numerical errors in approximating ${\mathcal L}_0^A\phi(x,t)$ 
minimized. It is straightforward that when the mesh size $h$ is fixed, 
the errors are minimized only when the length of the interval $[0, A]$ is the smallest.
Hence, we  choose $A = 2L$ in our scheme, leading to the integers $m = 1$ and $M = J$ 
in the scheme.
\end{remark}

%%%% %%%% %%%% %%%%
\begin{remark}\label{remark3-2} {\bf (Selection of the parameter $\gm$) } On the one hand, 
the selection of the parameter $\gm$ should ensure that the integral
 $\int_0^{\xi_1} 1/\xi^{-1+(\ap+\gm)}\ d\xi$ 
is convergent, equivalently,  we require $\gm <  (2-\ap)$. On the other hand,  to obtain a better accuracy 
from the trapezoidal type quadrature rule method,  we require that the positive constants  $\gm$ 
and $\sigma = 2-(\ap+\gm)$ as small as possible.  A simple calculation shows that 
$\gm = 1-\ap/2$ is the optimal constant to meet the above requirements.
 \end{remark}

% =======================================================================
%                                                          Section 4: 
% =======================================================================
\section{Fractional linear Schr\"{o}dinger equation}
\setcounter{equation}{0}
\label{section4}

In this section, we first show the difference between the standard and fractional Laplacian 
by giving one example and then numerically study the ground and first excited states of the fractional linear  
Schr\"{o}dinger equation in an infinite potential well.

% --------------------------------------------------------------------------------------------------------------------------
\subsection{Standard and fractional Laplacian}
\label{section2-3}

%%%% %%%% %%%% %%%% P1:
In the following, we use one example to show the difference between the standard Laplacian
$\Dt$ and the fractional Laplacian $-(-\Dt)^{\ap/2}$ defined  in (\ref{riesz}). Our example indirectly 
proves that the eigenfunctions of the standard Schr\"{o}dinger equation in an infinite potential well 
cannot be the eigenfunctions of the fractional Schr\"{o}dinger equation.

%%%% %%%% %%%% %%%% P2
Consider a function 
\bea\label{eq01}
u(x) = \left\{\begin{array}{ll}
\sin\left[\fl{\pi}{2}(1+x)\right], \ \  \ & x\in (-1,1),\\
0, & \mbox{otherwise},
\end{array}\right. \qquad x\in{\mathbb R},
\eea
which is continuous for $x\in{\mathbb R}$.  It is easy to compute
\beas
 -\Dt u(x) = \left\{\begin{array}{ll}
\fl{\pi^2}{4}\sin\left[\fl{\pi}{2}(1+x)\right], \ \  & x\in (-1,1),\\
0, & x\notin[-1, 1],
\end{array}\right. \quad x\in{\mathbb R},
\eeas
that is, except for $x = \pm 1$, the function $-\Dt u(x) = c u(x)$ with the constant  $c  = \pi^2/4$, 
which suggests  that $u(x)$ is an eigenfunction of the standard linear Schr\"{o}dinger equation 
in an infinite potential well with $c$  the corresponding eigenvalue. For the fractional cases, 
we plot the function $(-\Dt)^{\ap/2}u(x)$ in Figure \ref{F0}, 
where the results for $\ap < 2$
%******************************************************************
\begin{figure}[htb!]
\centerline{
\includegraphics[height=4.860cm,width=6.760cm]{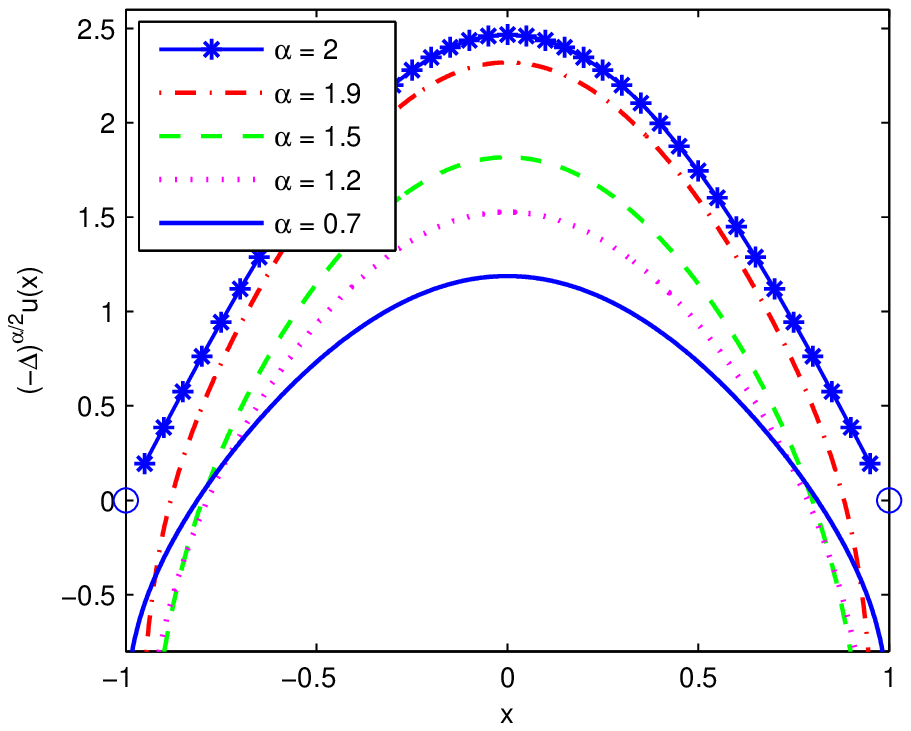}\hspace{-0.2cm}
\includegraphics[height=4.860cm,width=6.760cm]{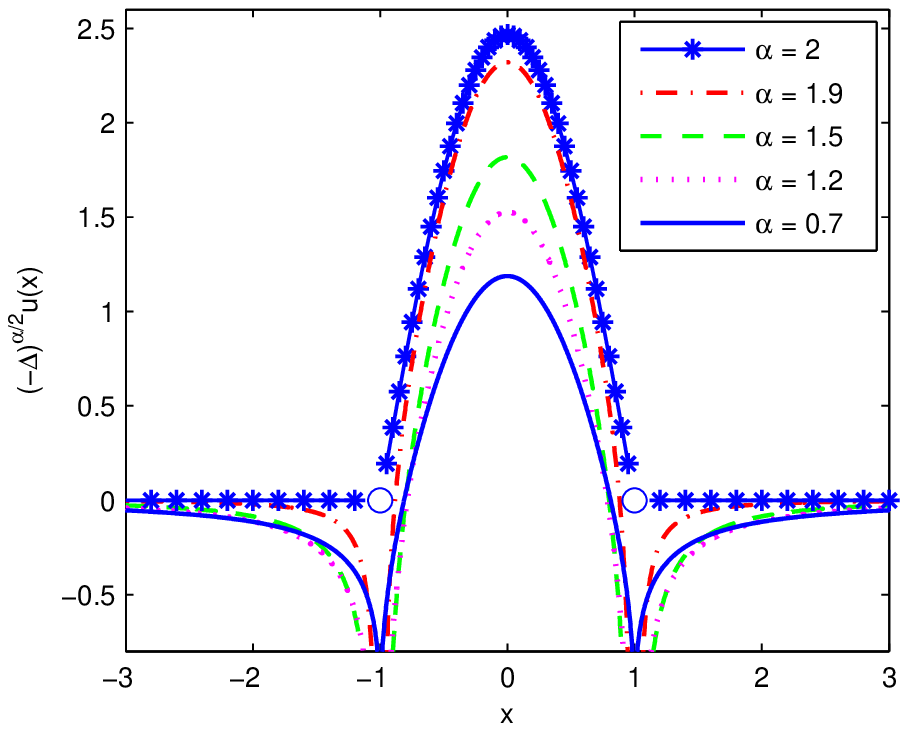}
}
\caption{The function $(-\Dt)^{\ap/2}u(x)$ for various $\ap$, where `o' represents the points 
where the function $(-\Dt)^{\ap/2}u$ does not exist.    The displayed domain: 
(a) $x\in (-1, 1)$; (b) $x\in(-3, 3)$. It shows that $u(x)$ in \eqref{eq01} is an eigenfunction of the 
standard linear Schr\"{o}dinger equation in an infinite potential well,  but it is not an eigenfunction 
of the fractional Schr\"{o}dinger equation. }\label{F0}\vskip -10pt
\end{figure}
%%%% %%%% %%%% %%%%
are computed  using the method proposed in Section \ref{section3}. For simplicity,  in Figure 
\ref{F0} we use $\ap = 2$ to represent the result of $-\Dt u$. 
 For $x\in(-1, 1)$,  Figure \ref{F0} (a) shows that  the function 
$-\Dt u(x)$ is always positive, and moreover it has the same shape as 
$u(x)$. However, if $\ap < 2$,  the function $(-\Dt)^{\ap/2}u(x)$ becomes negative near the 
boundaries $x = \pm 1$, and $(-\Dt)^{\ap/2}u(x) \neq cu(x)$ for any constant $c$.
Furthermore,  Figure \ref{F0} (b) shows that when $\ap < 2$,   the function 
$(-\Dt)^{\ap/2} u(x)$ is not always zero for $x\in{\mathbb R}\backslash[-1, 1]$,  albeit $u(x) \equiv 0$, which is 
completely different from the case of the standard Laplacian.
Since when $\ap < 2$  there is no nonzero constant $c$ satisfying 
$(-\Dt)^{\ap/2}u(x) = c u(x)$ for $x\in(-1, 1)$ or $x\in{\mathbb R}\backslash[-1, 1]$,  
$u(x)$ cannot be an eigenfunction of the fractional linear Schr\"{o}dinger equation in  an 
infinite potential well  \cite{Luchko2013, Herrmann2013}. 
For more discussions,  see Sections \ref{section4-1}. \\

%%%% %%%% %%%% %%%% P1:
In Sections \ref{section4-1}--\ref{section4-2}, the ground and first excited states of the fractional linear  
Schr\"{o}dinger equation in an infinite potential well are studied by numerically solving the 
fractional gradient flow in (\ref{GFDN})--(\ref{GFDN-BC}) with $\bt = 0$. 
In our simulations, we choose $L = 1$, equivalently, $\Og = (-1, 1)$ and 
$\Og_b = [-3, -1]\cup[1, 3]$.  The mesh size  is $h = 1/4096$,  and the time step is 
$\Dt t = 0.005$. The initial condition is chosen as 
\bea\label{phi0}
\varphi(x) = \sin\left[\fl{(s+1)\pi}{2}(1+x)\right],  \quad \ x\in\Og, \qquad s = 0\ \mbox{or}\  1,
\eea
where we choose $s = 0$ for computing the ground states, respectively, $s = 1$ for the first excited 
states.  We choose the tolerance $\varepsilon = 10^{-5}$ in (\ref{tol}).
In the following, we will use the subscripts ``\textit{g}"  and ``{\it 1}" to represent the ground states 
and the first excited states, respectively, and only the results inside of the infinite potential well 
(i.e.,  for $x\in\Og$) will be displayed,  since $\phi(x) \equiv 0$ for $x\notin\Og$. 

% ----------------------------------------------------------------------------------------------------------------------------
%                                                        Section 4-1:
% ----------------------------------------------------------------------------------------------------------------------------
\subsection{Ground states}
\label{section4-1}

%%%% %%%% %%%% %%%% P1:
Figure  \ref{F1} depicts  the wave function $\phi_g(x)$ of the ground states of the fractional 
linear Schr\"{o}dinger equation in an infinite potential well  for various  $\ap$.
%where the arrow indicates the change in the  wave function for progressively 
%increasing $\ap$. 
%******************************************************************
\begin{figure}[h]
\centerline{
(a)\includegraphics[height=5.160cm,width=7.460cm]{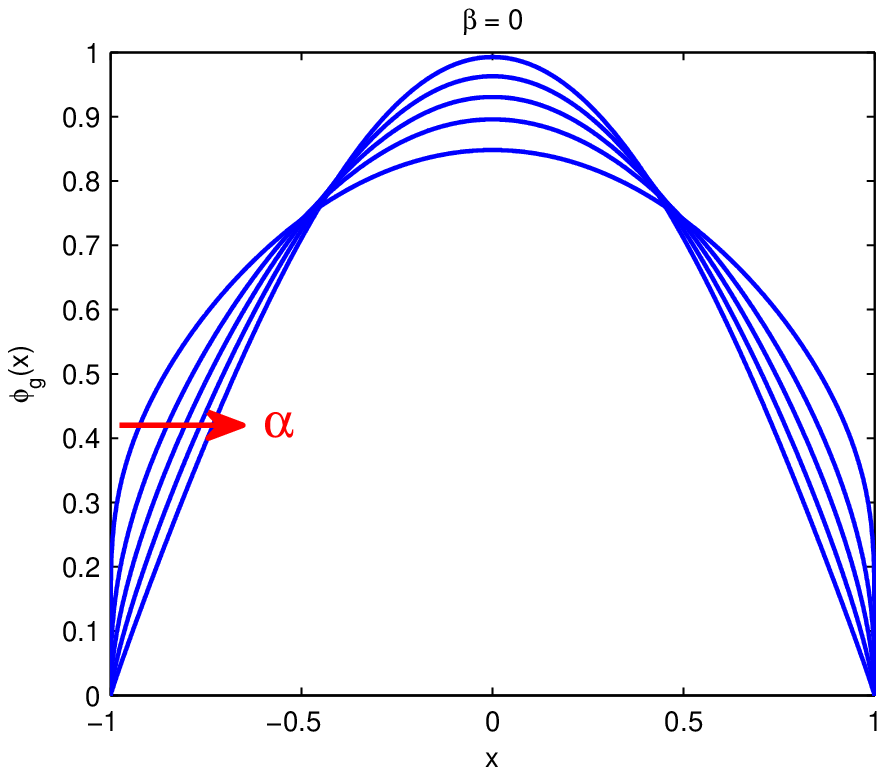}\hspace{-0.2cm}
(b)\includegraphics[height=5.10cm,width=4.860cm]{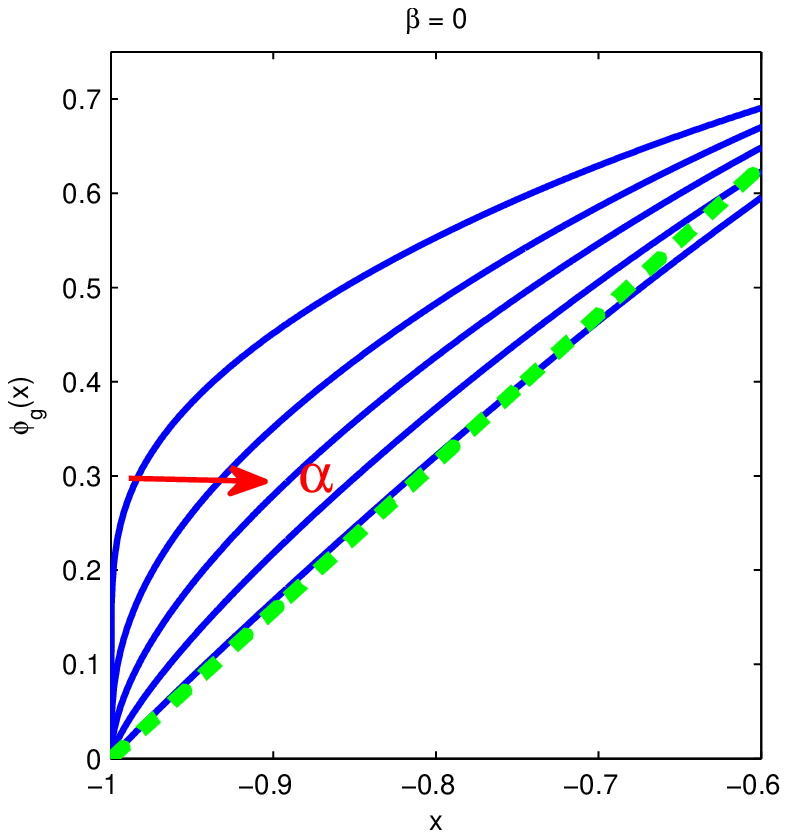}}
\caption{Ground states of the fractional linear ($\bt = 0$) Schr\"{o}dinger equation
 for $\ap = 0.2, 0.7, 1.1, 1.5$, and $1.9$, where the arrow indicates the change 
in the wave function  for progressively increasing $\ap$.  The plot  in (b) shows the change 
of the wave function near the boundary $x = -1$,  the green dotted line  $y=\fl{\pi}{2}(1+x)$ is 
presented for the sake of comparison.  The wave function $\phi_g(x)$ approaches $\sin[\pi(1+x)/2]$ 
as $\ap \to 2$, and its density near the boundaries is large if $\ap$ is small. 
 }\label{F1}\vskip -10pt
\end{figure}
%******************************************************************.
The wave function of the ground state is symmetric with respect to 
the center of the potential well $x = 0$, i.e., $\phi_g(x) = \phi_g(-x)$ for $x\in(-1, 1)$.   
The wave function monotonically increases for $x\in(-1, 0)$ and monotonically 
decreases for $x\in(0, 1)$, and it reaches the maximum value at $x = 0$.  Furthermore,  
the ground state of the fractional Schr\"{o}dinger equation in an infinite potential  well 
depends significantly on the parameter $\ap$.
If $\ap$ is small, the nonlocal interactions from the fractional Laplacian are strong, resulting in a 
flatter shape of the wave function. While if $\ap \to 2$, the wave function of the ground state converges to 
$\phi_g(x) = \sin\big[\pi(1+x)/2\big]$ -- the ground state solution of the standard Schr\"{o}dinger equation.  
In addition, Figure \ref{F1} (b) shows that the wave function changes quickly around $x = \pm 1$.  
The smaller the parameter $\ap$, the larger the density of the wave function near the boundaries. 

%%%% %%%% %%%% %%%%
Our observation  suggests that the eigenfunctions of 
the fractional  linear Shcr\"{o}dinger equation in an infinite potential well differ from those of 
the standard Schr\"{o}dingier equation, which confirms the conclusions made in 
\cite{Luchko2013, Herrmann2013}. Furthermore, our numerical results in Figure \ref{F1} 
are consistent with the ground states reported in \cite{Zoia2007}\footnote{See Figure 5 in 
\cite{Zoia2007} for absorbing boundary conditions. },  which were obtained from solving the 
eigenvectors of a large Toeplitz matrix representing the discrete fractional Laplacian.   
 However, our numerical method converges much faster.  Moreover, our method can be used to 
 compute the ground and first excited  states of  the fractional Schr\"{o}dinger equation not 
 only  in the linear cases  but also in the nonlinear cases.
 
 %%%% %%%% %%%% %%%% P2:
To further understand the properties of the ground states, we define the
 {\it expected value of position} for the $s$-th stationary state as
\bea
\langle x \rangle_s = \int_{\mathbb R} x\,|\phi_s(x)|^2 dx =  \int_{\Og} x\,|\phi_s(x)|^2 dx, 
\qquad s\in{\mathbb N}^0, 
\eea 
and the {\it variance in position} as
\bea\qquad 
{\rm Var}_s(x) = \int_{\mathbb R} \big(x-\langle x\rangle_s\big)^2\,|\phi_s(x)|^2 dx = 
\int_{\Og}\big(x-\langle x\rangle_s\big)^2\,|\phi_s(x)|^2 dx,\qquad s\in{\mathbb N}^0.
\eea
For the standard linear Schr\"{o}dinger equation,  the expected 
value of position of the $s$-th stationary state and its variance can be  exactly computed 
from \eqref{phis0}:
\bea\label{pos-var}
\langle x\rangle_s = 0\quad\ \  \mbox{and}\quad\ \  {\rm Var}_s(x) = 
\fl{L^2}{3}\left(1-\fl{6}{\pi^2(s+1)^2}\right), \qquad s\in{\mathbb N}^0,
\eea
that is,  for any $s\in{\mathbb N}^0$, the average position is always at $x = 0$ -- the 
center of the infinite potential well. The variance in position increases as $s$ increases, 
and as $s\to \infty$, the variance ${\rm Var}_s(x) \to L^2/3$.
While in the fractional cases,
%******************************************************************.
\begin{figure}[htb!]
\centerline{
\includegraphics[height=5.060cm,width=6.960cm]{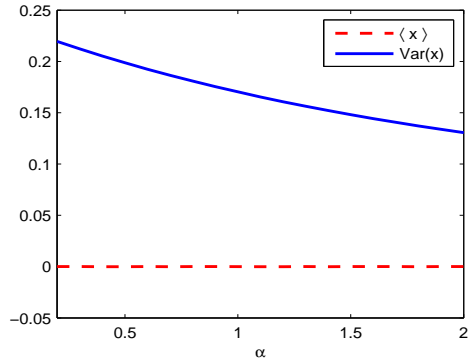}}\quad
\caption{The expected value of position and its variance of the ground state solutions of 
the fractional linear Schr\"{o}dinger equation. The expected value of 
position is zero for any $\ap\in(0, 2)$.
The smaller the parameter $\ap$, the stronger the nonlocal interactions, and thus 
the larger the scattering of particles, resulting in a larger variance in position. }\label{F1-1}\vskip -10pt
\end{figure}
%******************************************************************
Figure \ref{F1-1} shows the expected value of position in the ground state 
and its variance. Due to the symmetry of the wave function $\phi_g(x)$ with respect to 
$x = 0$, the expected value of the 
position $\langle x \rangle_g \equiv 0$, independent of the parameter $\ap$. However, the variance in 
position highly depends on the parameter $\ap$. The smaller the parameter $\ap$, 
the larger the variance in position. In fact, 
when $\ap$ decreases, the nonlocal interactions represented by the fractional Laplacian become
stronger, resulting in a larger scattering of particles. Hence, the decrease in the parameter $\ap$ 
leads to an increase in the variance in  position.

%%%% %%%% %%%% %%%% Table 1
In Table \ref{T1},  we present our simulated eigenvalues $\mu_g^h$ of 
the ground states  and compare them with the lower and upper bound estimates in 
\cite{Banuelos2004} and the asymptotic approximation $\mu_g^a$ in \cite{
Kwasnicki2012}. 
Note that in the linear case (i.e., $\bt = 0$), the eigenvalue 
$\mu_s$ is equal to the energy $E_s$ for any $\ap \in(0, 2]$ and $s\in{\mathbb N}^0$. 
From Table \ref{T1}, we find that the eigenvalue  $\mu_g$ increases as $\ap\geq 0.3$ increases,  
and as $\ap \to 2$, it converges to $\mu_g = \pi^2/4$ --
the eigenvalue of the ground states of the standard linear Schr\"{o}dinger equation.
%******************************************************************
\begin{table}[ht!]
\begin{center}
\begin{tabular}{|c||c|c|c|c|c|}
\hline
$\ap$ & $\mu_{g}^{\rm lower}$ in \cite{Banuelos2004}&$\mu_g^h$ & $\mu_{g}^{\rm upper}$in 
\cite{Banuelos2004}& $\mu_g^{a}$ 
in \cite{Kwasnicki2012}& $|\mu_g^h - \mu_g^{a}|$\\
\hline
0.1 &  0.9514 &0.9726 & 0.9786 &0.9809 &0.0083\\
0.2 & 0.9182 &0.9575 & 0.9675 & 0.9712 & 0.0137\\
0.3 &  0.8975 &0.9528& 0.9655 &0.9699 & 0.0172     \\
0.5 & 0.8862 & 0.9702& 0.9862  &0.9908&0.0206\\
0.7 & 0.9086 &1.0203&1.0383 &1.0418 &  0.0215\\
0.9 & 0.9618 &1.1032& 1.1227&1.1241 & 0.0209\\
1    &1 & 1.1578 & $3\pi/8$ &$3\pi/8$ &0.0203\\
1.1 & 1.0465 &1.2222&1.2432&1.2415 &  0.0194\\
1.3 & 1.1667& 1.3837&1.4064&1.4007 &  0.0170 \\
1.5 & 1.3293 &1.5976&1.6223 &1.6114 &  0.0138\\
1.7 & 1.5447 &1.8779& 1.9053 & 1.8873 &  0.0094\\
1.9 & 1.8274 &2.2441& 2.2747 &2.2477 & 0.0036\\
1.99 & 1.9817 &2.4437& 2.4761&2.4441 &  0.0004\\
\hline
\end{tabular}
\caption{The eigenvalue $\mu_g$ of the ground states of the 
fractional linear ($\bt = 0$) Schr\"{o}dinger equation in an infinite potential well. 
$\mu^h_g$ represents our numerical results, 
$\mu_{g}^{a}= \left(\fl{\pi}{2}-\fl{(2-\ap)\pi}{8}\right)^\ap$ is the asymptotic approximation  obtained 
in \cite{Kwasnicki2012}, and $\mu_{g}^{\rm lower}$ and $\mu_{g}^{\rm upper}$ are the lower and upper 
bounds estimated in \cite{Banuelos2004}, respectively. It shows that our numerical results $\mu_g^h$ are consistent with the analytical estimates 
provided in \cite{Banuelos2004, Kwasnicki2012}, and as $\ap \to 2$,  the eigenvalue 
$\mu_g\to \pi^2/4$ -- the eigenvalue of the ground states of the standard linear Schr\"{o}dinger equation.}\label{T1} 
\end{center}\vskip -10pt
\end{table}
%******************************************************************
Our numerical results are consistent with the lower and upper bounds estimates of the 
eigenvalues provided in \cite{Banuelos2004, Deblassie2004, Chen2005}, i.e., $\mu_g^h 
\in [\mu_g^{\rm lower}, \mu_g^{\rm upper}]$ for any $\ap\in(0, 2)$. 
Moreover,  our results suggest that for the eigenvalues of the ground states,  the lower and upper bounds 
 obtained in \cite{Banuelos2004} are much sharper than those in \cite{Deblassie2004, Chen2005}. 
 In addition, we compare our numerical results with the asymptotic approximation 
$\mu_g^a = [\pi/2 - (2-\ap)\pi/8]^\ap$ obtained in \cite{Kwasnicki2012} and find that the asymptotic 
results are more accurate as $\ap \to 2$. 

%Analytically, two different results on the eigenvalues $\mu_g$ have been reported in the literature 
%\cite{Laskin2000, Laskin2002, Kwasnicki2012, Dong2013}.  On the one hand,  Laskin
%and many others  claimed that the eigenvalues of the fractional linear Schr\"{o}dinger 
%equation in infinite potential well is given by $\mu_s = \big[(s+1)\pi/2\big]^\ap$, as a modification 
%of the eigenvalue $\mu_s =\big[(s+1)\pi/2\big]^2$  for the standard Schr\"{o}dinger equation 
%\cite{Laskin2000, Laskin2002, Dong2013}. It implies that when $s = 0$, the eigenvalue of 
%the ground state is $\mu_g = (\pi/2)^\ap$. However, our numerical simulations show that 
%their results are incorrect. 

% ================================================================================
\subsection{The first excited states}
\label{section4-2}

%%%% %%%% %%%% %%%% P1:
Figure \ref{F2} depicts the wave function $\phi_1(x)$ of the first excited states of the fractional linear 
 Schr\"{o}dinger equation in an infinite potential well for various $\ap$. 
 It shows that the wave function $\phi_1(x)$ varies for different values of $\ap$, and as $\ap \to 2$,  
it converges to $\phi_1(x) =  \sin(\pi(1+x))$ -- the first excited state solution of the standard  linear 
%******************************************************************
\begin{figure}[ht!]
\centerline{
(a)\includegraphics[height=5.26cm,width=7.46cm]{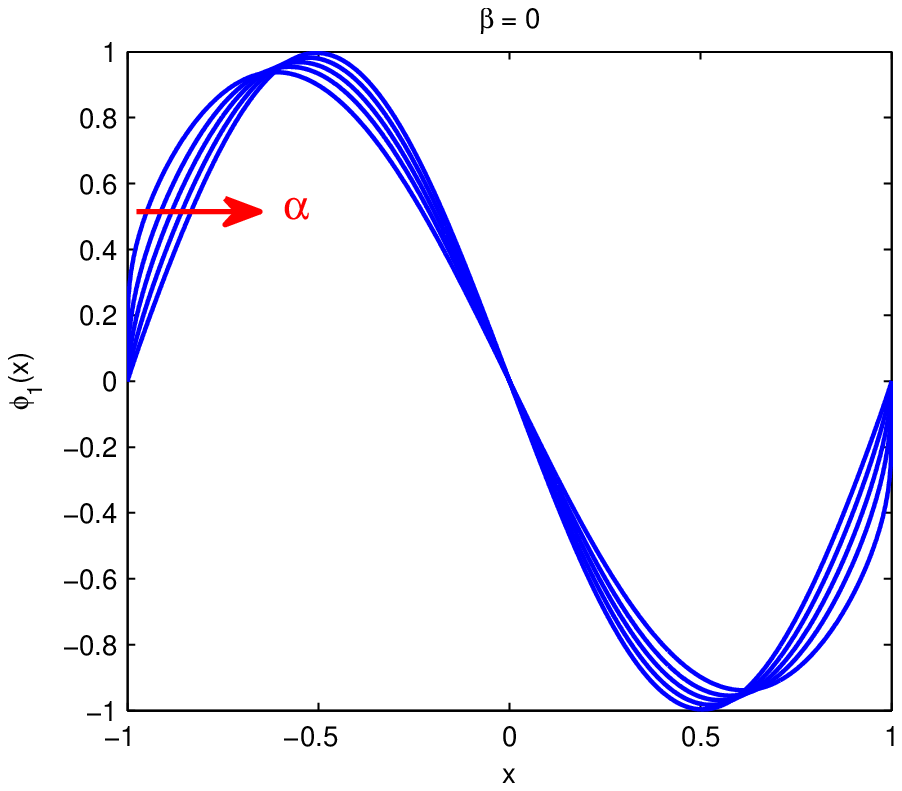}\hspace{-0.2cm}
(b)\includegraphics[height=5.2cm,width=4.860cm]{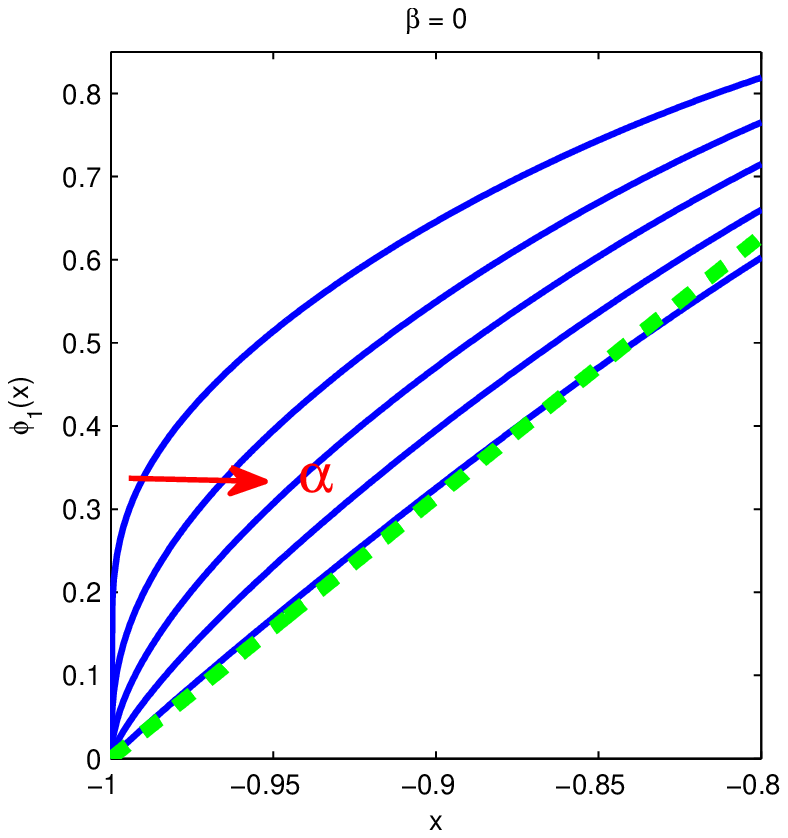}
}
\caption{The first excited states of the fractional linear  Schr\"{o}dinger equation 
for $\ap = 0.2, 0.7, 1.1, 1.5$, and $1.9$, where the arrow indicates the change 
in the wave function for progressively 
increasing $\ap$.  The plot in (b) shows  the change of the wave function near the boundary $x = -1$, the 
green dotted line $y=\pi(1+x)$ is presented for the sake of comparison.  The wave function $\phi_1(x)$
approaches $\sin[\pi(1+x)]$ as $\ap \to 2$, and its density near the boundaries is large if $\ap$ is small.  
}\label{F2}\vskip -10pt
\end{figure}
%******************************************************************
%where the arrow indicates the change in the wave function for progressively increasing $\ap$.
Schr\"{o}dinger equation in an infinite potential well.  In addition, the wave function  $\phi_1(x) $ 
of the first excited states is antisymmetric with respect  to the center of the infinite potential well $x = 0$, 
i.e., $\phi_1(x) = -\phi_1(-x)$ for $x\in(-1,1)$  and $\phi(0) = 0$. 
For the standard linear Schr\"{o}dinger equation, 
 $\phi_1(x)$ is  also symmetric on each subinterval 
$(-1, 0)$ and $(0, 1)$, but in the fractional cases the wave function 
loses the symmetry in subintervals. 

%%%% %%%% %%%% %%%% P2:
Denote $\rho_1(x) = |\phi_1(x)|^2$  as the position density  of the first excited states.  
The fact that  $\phi_1(x)$ is antisymmetric about the center of the potential well implies that the 
position density $\rho_1(x)$ is symmetric with respect to $x = 0$. 
Furthermore, there exist two points ${x}_{c}\in(0, 1)$ and $-x_c\in(-1, 0)$ at 
which the density function $\rho_1(x)$ reaches its maximum values, 
%******************************************************************.
\begin{figure}[htb!]
\centerline{
\includegraphics[height=5.060cm,width=6.760cm]{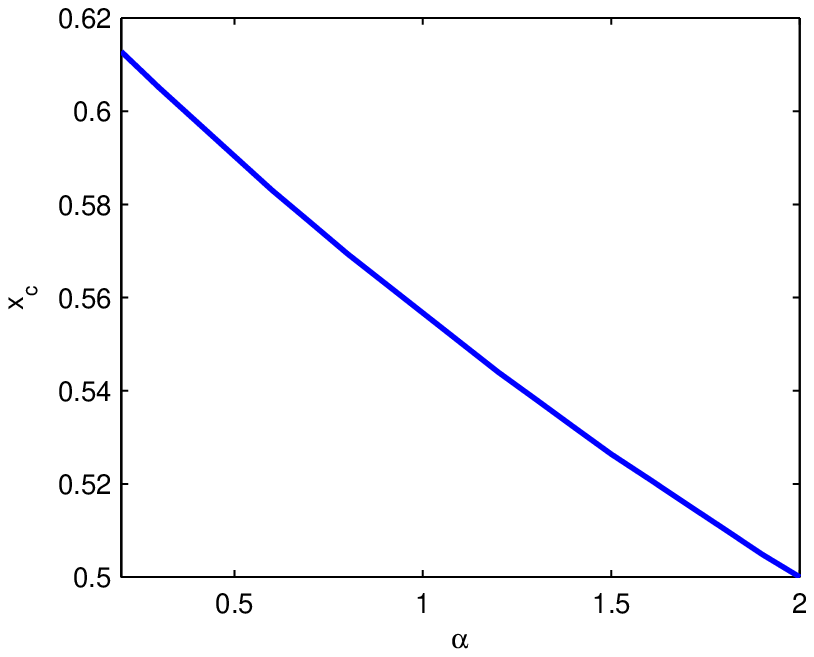}\hspace{-0.2cm}
\includegraphics[height=5.060cm,width=6.760cm]{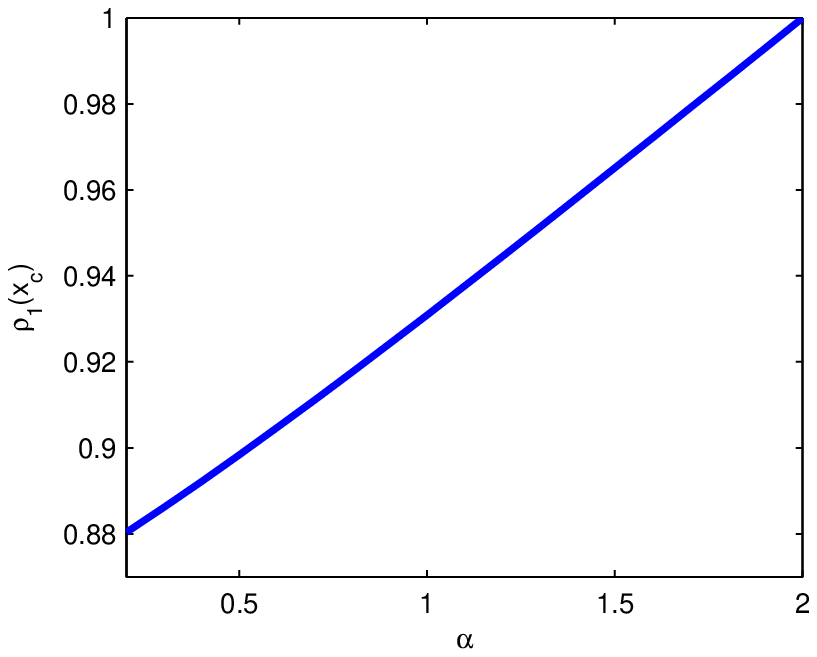}}\quad
\caption{(a) The point $x_c$  versus the parameter 
$\ap$.  (b) The maximum value  $\rho_1(x_c)$ versus the parameter $\ap$.}\label{F2-1}
\vskip -10pt
\end{figure}
%******************************************************************
i.e., $\rho_1(x_c) = \rho_1(-x_c) = \max_{x\in[-1,1]} \{\rho_1(x)\}$.  
The point $x_c$ varies for different parameter $\ap$. Figure \ref{F2-1} shows the 
values of $x_c$ and $\rho_1(x_c)$ for various $\ap$. We see that the larger 
the parameter $\ap$,  the smaller the value of $x_c$, but 
the larger the density function $\rho_1(x_c)$, 
%  For the purpose of easy comparison, 
%we present a dashed linear line representing {\color{red}$y = 0.0344\ap+0.9313$}
 %in  Figure \ref{F2-1} (b),  
and the maximum value 
$\rho_1(x_c)$ increases almost linearly as the parameter $\ap$.
In particular,  in the first excited states of the standard linear Schr\"{o}dinger equation, 
the point $x_c = \fl{1}{2}$  and the maximum density function  $\rho_1(\pm \fl{1}{2}) = 1$.
 
%%%% %%%% %%%% %%%%
In addition, Figure \ref{F2-2}  shows the expected value of position and its  variance 
in the first excited states.  For any $\ap \in(0, 2)$, the expected value 
$\langle x \rangle_1 \equiv 0$, due to the antisymmetry of the wave function $\phi_1(x)$, 
while the variance in position changes for different $\ap$.
%******************************************************************.
\begin{figure}[htb!]
\centerline{
\includegraphics[height=5.060cm,width=6.960cm]{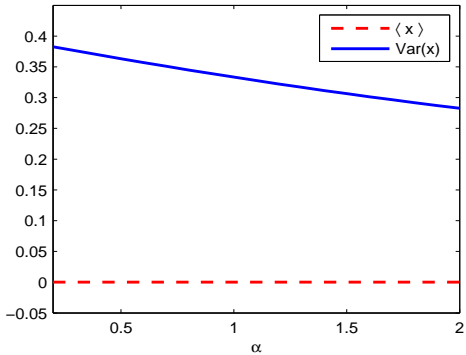}}
\caption{The expected value of position and its variance of the first excited 
states of the fractional linear Schr\"{o}dinger equation.  The expected 
value of position is zero for any $\ap\in(0, 2)$.
The smaller the parameter $\ap$, the stronger the nonlocal interactions, and 
thus  the larger the scattering of particles, resulting in a larger variance in position. }\label{F2-2}
\vskip -10pt
\end{figure}
%******************************************************************
The smaller the parameter $\ap$,  the stronger 
the scattering of particles, and thus the larger the variance in position, 
which is similar to our observation in Figure \ref{F1-1} for the ground states. 
However, we find that  for a fixed $\ap$, the variance ${\rm Var}_1(x) > {\rm Var}_g(x)$.  In fact, the energy 
of the first excited states is higher than that of the ground states,  and consequently the scattering 
of particles in the first excited states is stronger, which leads to a larger variance 
of the first excited states.

%%%% %%%% %%%% %%%%
Table \ref{T2} shows the eigenvalues of the first excited states, 
where $\mu_1^h$ represents  our numerical results,  $\mu_1^a =  
\left[\pi - {(2-\ap)\pi}/{8}\right]^\ap$ is the asymptotic approximation reported  
in \cite{Kwasnicki2012}, and $\mu_1^{\rm lower}$ and $\mu_1^{\rm upper}$ are the lower 
and upper bound estimates provided in \cite{Chen2005,Deblassie2004}, respectively.
We find that the eigenvalue of the first excited states increases as $\ap$ increases, and as $\ap \to 2$,  
it converges to $\mu_1 = \pi^2$ -- the eigenvalue of the first excited states of the standard 
linear Schr\"{o}dinger equation in an infinite potential well.
%Our numerical results again suggest that the eigenvalues $\mu_1 = \pi^\ap$ reported in \cite{Laskin2000, 
%Laskin2000-1, Laskin2000-2, Laskin2002, Bayin2012, Dong2013} for fractional 
%Schr\"{o}dinger equation are incorrect, 
Our numerical results $\mu_1^h$ are consistent with the estimates obtained in 
\cite{Deblassie2004, Chen2005} as well as the asymptotical approximations $\mu_1^a$ 
reported in \cite{Kwasnicki2012}. Furthermore, our results suggest that 
the asymptotic results in \cite{Kwasnicki2012} are more accurate for the 
first excited states than for the ground states, as the asymptotic approximation $\mu_1^a$ has the 
error  $O\left((2-\ap)/(s+1)\sqrt{\ap}\right)$ for $s\in{\mathbb N}^0$.  

%******************************************************************
\begin{table}[ht!]
\begin{center}
\begin{tabular}{|c||c|c|c|c|c|}
\hline
$\ap$ &$\mu_1^{\rm lower}$ in \cite{Chen2005,Deblassie2004}  &$\mu_1^h$&  $\mu_1^{\rm upper}$ in \cite{Chen2005,Deblassie2004} &$\mu_1^{a}$ 
in \cite{Kwasnicki2012}& $|\mu_1^h - \mu_1^{a}|$\\
\hline
0.1 & 0.5606&1.0922 & 1.1213 & 1.0913 & 0.0009 \\
0.2 & 0.6286 & 1.1966& 1.2573 & 1.1948&0.0018\\
0.3 & 0.7049 &1.3148 &1.4098 &  1.3132 & 0.0026\\
0.5 &0.8862  &1.6016  & 1.7725  &1.5977& 0.0039\\
0.7 &1.1142  &1.9733 & 2.2285 &1.9683& 0.0050\\
0.9 & 1.4009  &2.4583 & 2.8018 &2.4526 & 0.0057\\
1 & $\pi/2$ & 2.7549 & $\pi$ & $7\pi/8$ &0.0060\\
1.1 & 1.7613&3.0954 &  3.5226&3.0892& 0.0062\\
1.3 & 2.2144   &3.9380  & 4.4289  &3.9319&0.0061\\
1.5 &2.7842 &5.0600  &5.5683 &5.0545 & 0.0055\\
1.7 & 3.5005 &6.5646  &  7.0009&6.5605&0.0041\\
1.9 &  4.4010&8.5959 &  8.8021  &8.5942&0.0017\\
1.99 & 4.8786 &9.7332 & 9.7573& 9.7330&0.0002\\
\hline
\end{tabular}
\caption{The eigenvalue $\mu_1$ of the first excited states of the fractional linear 
Schr\"{o}dinger equation in an infinite potential well.  $\mu^h_1$ represents our numerical results, 
$\mu_{1}^{a}= \left(\pi-\fl{(2-\ap)\pi}{8}\right)^\ap$ is the asymptotic approximation  
obtained in \cite{Kwasnicki2012}, and $\mu_1^{\rm lower} = \pi^\ap/2$ and $\mu_1^{\rm upper}
= \pi^\ap$ are 
the lower and upper bounds estimated in \cite{Chen2005,Deblassie2004}, respectively. 
 It shows that our numerical results $\mu_1^h$ are consistent with the analytical estimates 
obtained in \cite{Chen2005,Deblassie2004, Kwasnicki2012},  and as $\ap \to 2$,  
the eigenvalue $\mu_1\to \pi^2$ -- the eigenvalue of the 
 first excited states of the standard  linear Schr\"{o}dinger equation.  }\label{T2}
\end{center} \vskip -10pt
\end{table}
%******************************************************************

%\begin{remark}
%Our results in this paper can be applied to explain the ground states and the first excited 
%states of the fractional Schr\"{o}dinger equation in the unit disk. The wave function is equivalent 
%to that in a diameter.  While the eigenvalues in the unit disk can be obtained by multiplying $\pi$ 
%to our eigenvalues.
%\end{remark}

% =======================================================================
%                                                             Section 5:
% =======================================================================
\section{Fractional nonlinear Schr\"{o}dinger equation}
\label{section5}

%%%% %%%% %%%% %%%%
There have been many discussions on the stationary states (or eigenfunctions) of the fractional 
linear Schr\"{o}dinger equation in an infinite potential well based on different representations of 
the fractional Laplacian $-(-\Dt)^{\ap/2}$. However, to the best of our 
knowledge, no study has been reported  in the nonlinear  ($\bt \neq 0$) cases yet. 
In this section, we numerically study the ground and  first excited states of the fractional 
Schr\"{o}dinger equation with repulsive nonlinear interactions (i.e., $\bt > 0$) and attempt to 
understand  the effects of the local (or short-range) interactions and the competition of the local and 
nonlocal interactions.  
%%%% %%%% %%%% %%%% 
In our simulations, we choose  $L = 1$, the mesh 
size $h = 1/4096$,  the time step $\Dt t = 0.005$, and the convergence tolerance 
$\varepsilon = 10^{-5}$ in \eqref{tol}. The initial condition is chosen as defined  in (\ref{phi0}). 

% =======================================================================
\subsection{Ground states}
\label{section5-1}

%%%% %%%% %%%% %%%%
Figure \ref{F3} shows the wave function $\phi_g(x)$ of the ground states of the
 fractional nonlinear Schr\"{o}dinger equation in an infinite potential well for various $\ap$ 
and $\bt$.
% where the arrow indicates the change  in wave functions for progressively 
%increasing $\ap$.
% *********************************************************************
\begin{figure}[ht!]
\centerline{
\includegraphics[height=4.8600cm,width=6.660cm]{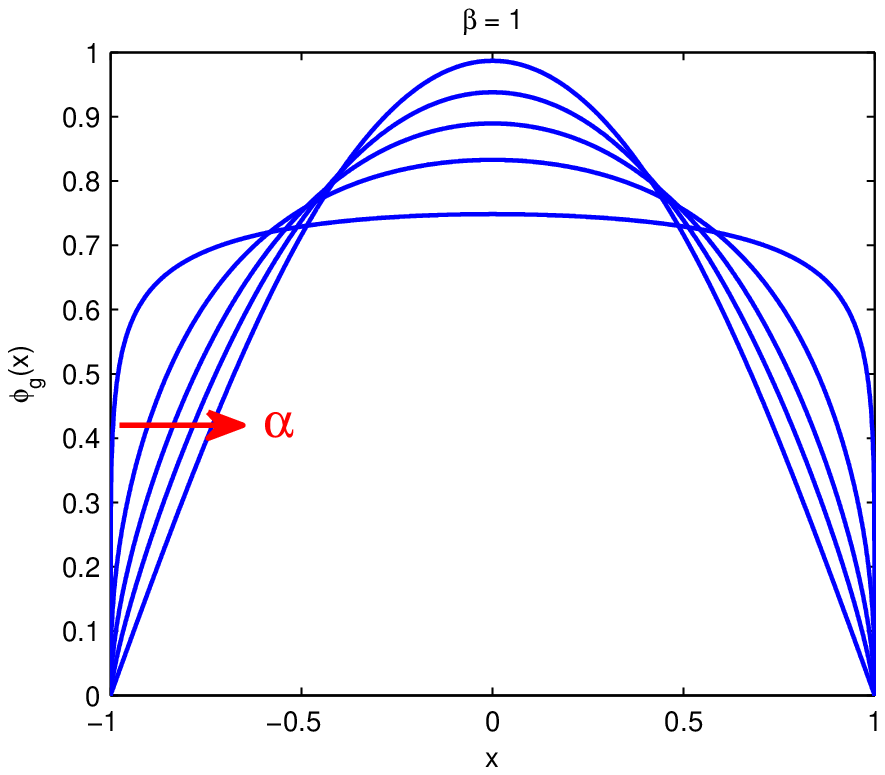}
\includegraphics[height=4.8600cm,width=6.660cm]{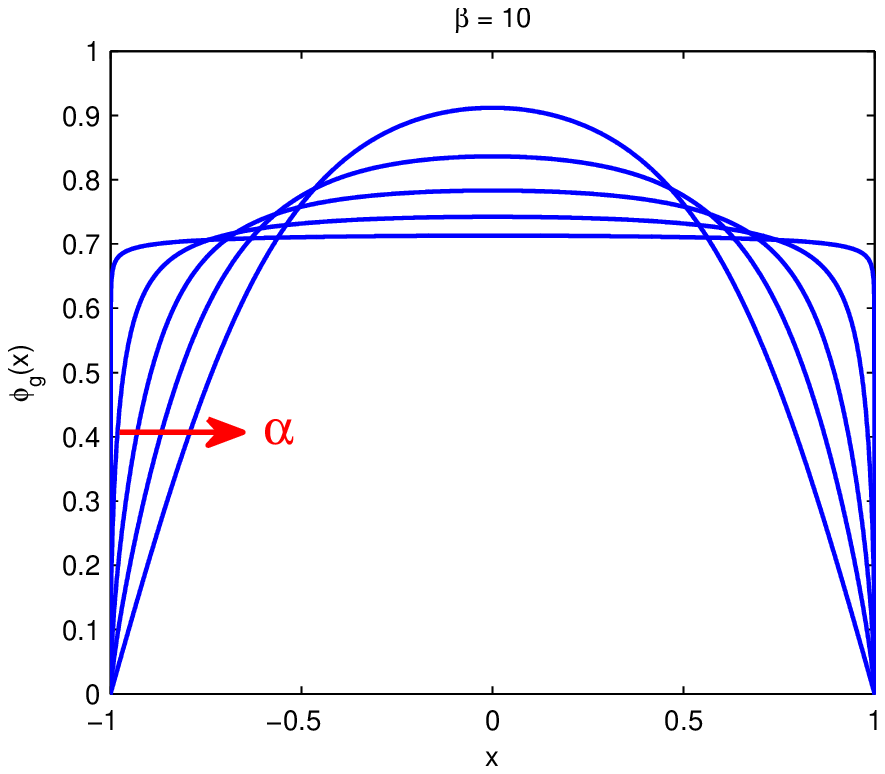}
}
\centerline{
\includegraphics[height=4.8600cm,width=6.660cm]{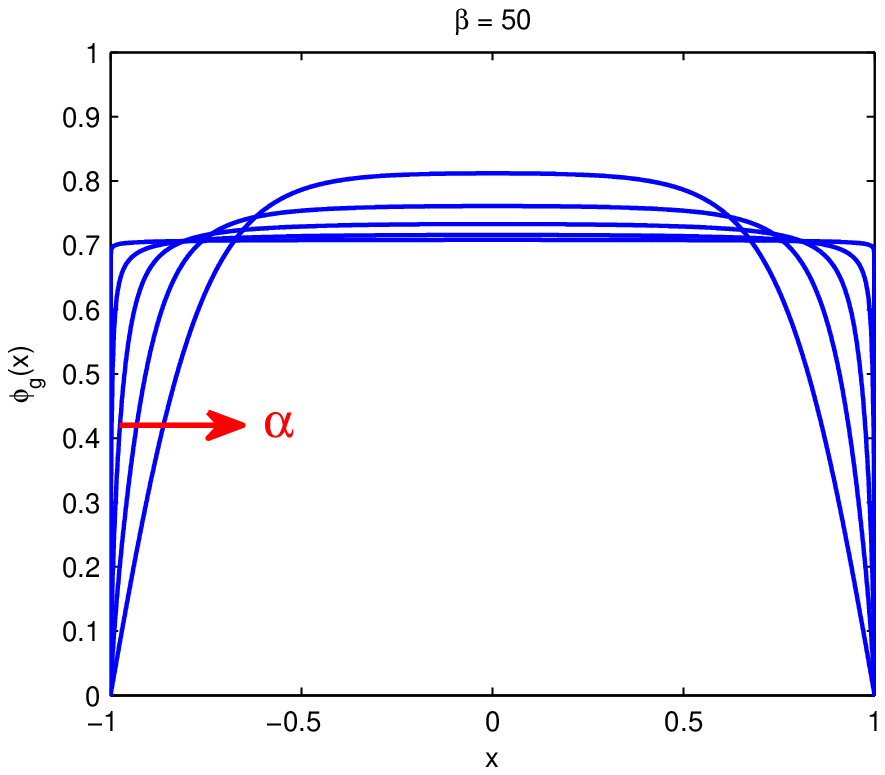}
\includegraphics[height=4.8600cm,width=6.660cm]{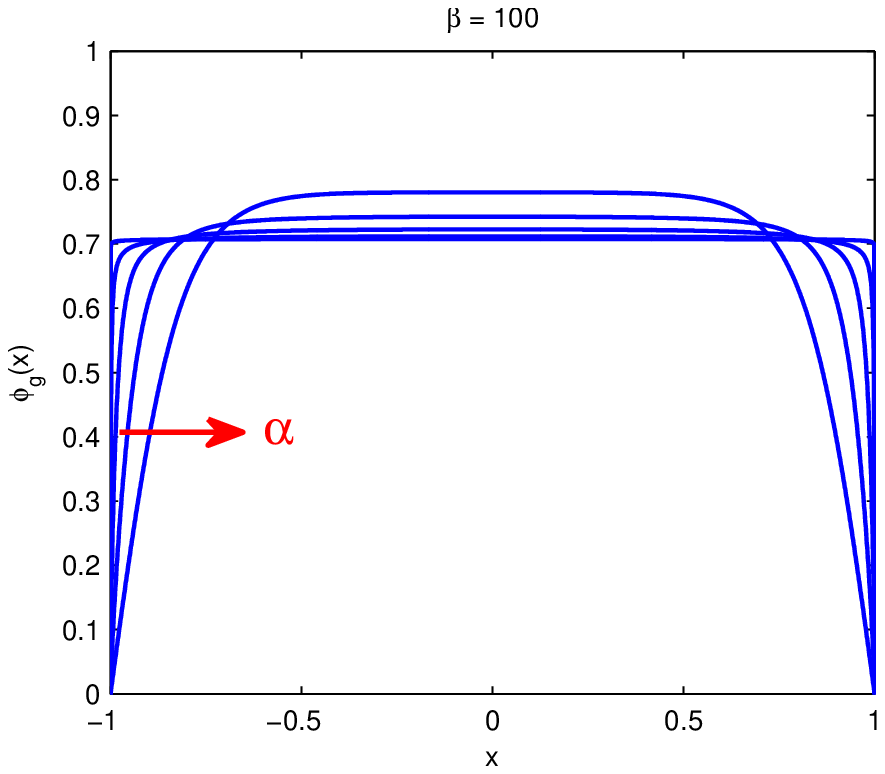}
}
\caption{Ground states of the fractional nonlinear Schr\"{o}dinger equation
 for $\ap = 0.2, 0.7, 1.1, 1.5$, and $1.99$, where the arrow indicates the 
 change in the wave function for progressively increasing $\ap$.  
 Boundary layers exist if $\ap$ is small or $\bt$ is large. }\label{F3}\vskip -10pt
\end{figure}
% *********************************************************************
The wave function of the ground states $\phi_g(x)$ is always symmetric with respect to 
the center of the infinite potential well $x = 0$.
As $\ap \to 2$, the wave function converges to the ground state solution of the 
standard Schr\"{o}dinger equation  with the same nonlinear parameter 
$\bt$. In contrast to the linear cases,  
the local repulsive interactions may lead to  boundary layers in the ground states.
Here, we divide our discussions into two  interaction regimes:   
the weak interactions when $\bt \sim o(1)$ and the strong interactions when $\bt \gg 1$. 
For $\bt \sim o(1)$, the effects of local repulsive interactions are significant 
only when $\ap$ is small, resulting in two boundary layers at $x = \pm 1$ (cf.  the case of 
$\ap = 0.2$ and $\bt = 1$ in Figure \ref{F3}).
While in the strongly interacting cases (e.g. $\bt = 50$ or $100$), the local interactions 
become  significant for any $\ap\in(0, 2)$.  
Due to the normalization condition,  the wave function $\phi_g(x)$ inside the potential well 
tends to approach the value $\sqrt{2}/2$. 
However,  since the wave function $\phi_g(x) \equiv 0$ at $x = \pm1$, 
two layers emerge at the boundaries of the potential well.

%%%% %%%% %%%% %%%% P2:
In addition, Figure \ref{F3} shows that the width of the boundary layers depends on both 
$\ap$ and $\bt$, and either increasing $\bt$ or decreasing $\ap$ can lead to the thinner boundary 
layers. That is, with the presence of the local interactions,  the strong local or 
nonlocal interactions can cause a sharp change in wave functions 
near the boundaries. To further understand this phenomenon,  
we define $w$ as the width of the boundary layers in the ground states.   
%******************************************************************
\begin{figure}[ht!]
\centerline{
(a)\includegraphics[height=5.060cm,width=6.660cm]{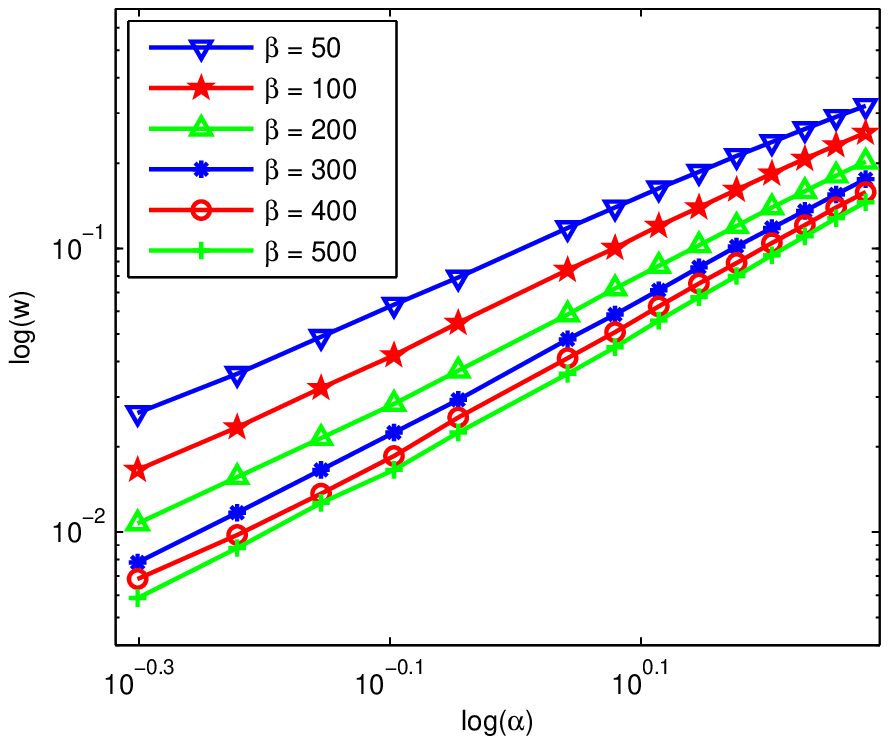}\hspace{-0.6cm}
(b)\includegraphics[height=5.060cm,width=6.660cm]{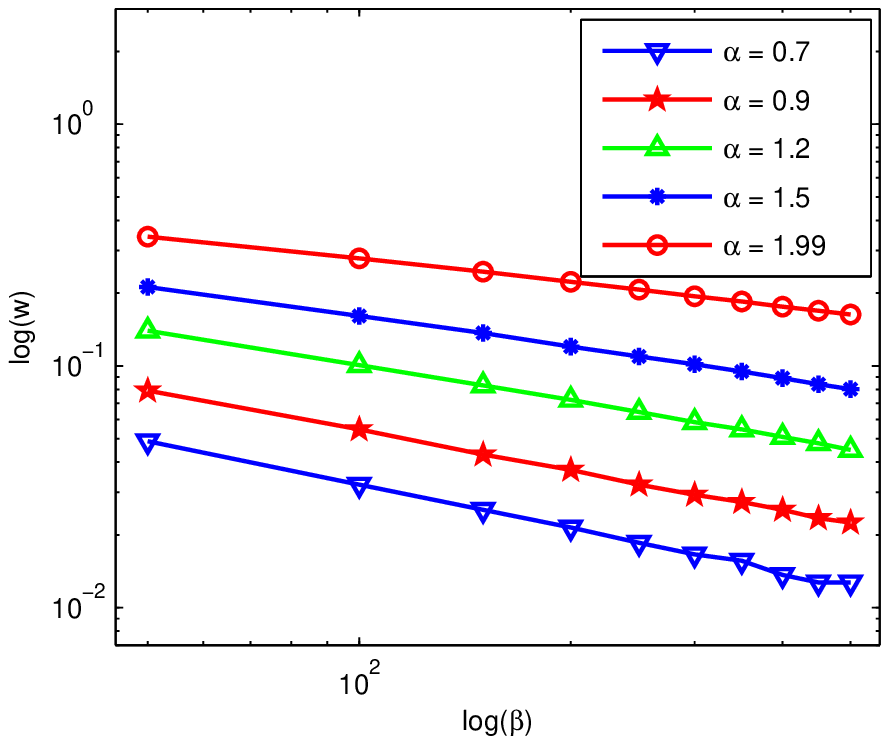}}
\caption{Log-log plots of:  (a) the width of boundary layers versus $\ap$, and (b) 
the width of boundary layers  versus $\bt$,  in the ground states of the fractional nonlinear 
Schr\"{o}dinger equation in  an infinite potential well. It suggests that the width of boundary layers 
$w$ increases (resp. decreases) 
as a power function of $\ap$ (resp. $\bt$).}\label{F3-1}
\end{figure}
%******************************************************************
In our simulations,  it is computed by $w = 1-|\bar{x}|$, where
$\bar{x}$ satisfies 
\beas  
\left|\fl{\p\phi_g(x)}{\p x}\Big|_{x = \bar{x}} \right|= \eta,\qquad \mbox{with $\eta\sim O(1)$ a constant}.
\eeas Here, we choose $\eta =  {\sqrt{2}}/{2}$. 
In Figure \ref{F3-1}, we present the log-log plots of the width of the boundary layers $w$ versus the parameter $\ap$ and $\bt$.
On the one hand, Figure \ref{F3-1} (a) shows that when $\bt$ is fixed, $\log(w)$ linearly increases as 
$\log(\ap)$ increases,  and thus we have
\bea\label{wa}
w =  c\ap^{p}, \qquad \mbox{where $c$ and $p>0$ are constants.}
\eea
The constant $p$ in (\ref{wa}) may vary for different parameter $\bt$.
On the other hand, Figure \ref{F3-1} (b) shows that when $\ap$ is fixed, 
$\log(w)$ linearly decreases as $\log(\bt)$  increases, and 
\bea\label{wb}
w = \fl{d}{\bt^{q}}, \qquad \mbox{where $d$ and $q>0$ are constants.}
\eea
The constant $q$ in (\ref{wa}) may be slightly different for various parameter $\ap$.
In particular,  the width of the boundary layers  $w\sim O\left(1/\sqrt{\bt}\right)$, 
equivalently, $q =1/2$ in (\ref{wb}) for the standard Schr\"{o}dinger equation \cite{Bao2007}.

%%%% %%%% %%%% %%%%
Since the wave function $\phi_g(x)$ is symmetric with respect to $x = 0$, 
the expected value of  position $\langle x \rangle_g\equiv 0$, 
independent of the parameters $\ap$ and $\bt$. Here, we omit showing it for brevity.
Figure \ref{F4} depicts the variance in position of the ground states of the fractional nonlinear 
Schr\"{o}dinge equation for various parameters $\ap$ and $\bt$. 
It shows that the variance in position monotonically increases as $\ap$ decreases or $\bt$ increases, 
implying that the strong local or nonlocal interactions yield a large scattering of particles 
in the potential well.  
%******************************************************************
\begin{figure}[ht!]
\centerline{
\includegraphics[height=5.06cm,width=6.960cm]{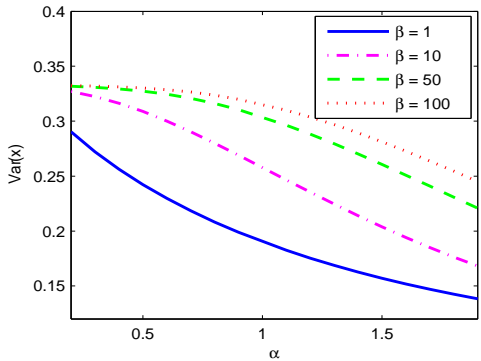}}
\caption{The variance in position of the ground states of the fractional nonlinear Schr\"{o}dinger 
equation in an infinite potential well. It increases as the local interaction 
parameter $\bt$ increases, but decreases as the nonlocal interaction parameter $\ap$ increases. The variance converges to $1/3$ as $\ap \to 0$ and $\bt \to \infty$. }\label{F4}
\end{figure}
%******************************************************************
Comparing Figure \ref{F4} to Figure \ref{F1-1}, we find that a small local interaction (e.g., $\bt = 1$) 
can significantly change the variance in position when $\ap$ is small.
In addition,  Figure \ref{F4} suggests that when $\bt$ is small (e.g., $\bt = 1$), 
the nonlocal interactions from the fractional Laplacian are dominant, and the variance decreases 
concave up  as $\ap$ increases, similar to the cases of $\bt = 0$.
When $\bt$ is large, the local repulsive interactions become dominant, and the decrease of the 
variance becomes concave down as $\ap$ increases.
In addition, due to the constraint $\|\phi_g\|^2 = 1$,  the variance of  position converges to 
 $1/3$ as $\ap \to 0$ or $\bt \to \infty$. 

%%%% %%%% %%%% %%%% 
In Table \ref{T3}, we present the simulated eigenvalue $\mu_g^h$ and the kinetic energy $\mu_{g,\rm kin}^h$ 
of the ground states of the fractional nonlinear Schr\"{o}dinger equation in an infinite potential 
well, where the kinetic energy $\mu_{g, \rm kin}$ and the interaction energy $\mu_{g, \rm int}$ 
of the $s$-th stationary state are defined by
\beas
\mu_{s, \rm kin} =  \int_{\mathbb R}\phi_s^*(-\Dt)^{\ap/2}\phi_s\, dx, \qquad 
\mu_{s, \rm int} =  \bt\int_{\mathbb R}|\phi_s(x)|^2 dx, \qquad s\in{\mathbb N}^0,
\eeas
respectively. It is easy to see that the eigenvalue $\mu_s = \mu_{s,\rm kin} + \mu_{s, \rm int}$, 
and when $\bt = 0$,  we have $\mu_s \equiv \mu_{s, \rm kin}$ for any $\ap \in (0, 2]$.
%******************************************************************
\begin{table}[htb!]
\begin{center}
\begin{tabular}{|c||c|c||c|c||c|c||c|c|}
\hline
\multirow{2}{*}{$\ap$}&  \multicolumn{2}{c||}{$\bt = 1$} &  \multicolumn{2}{c||}{$\bt = 5$} & 
 \multicolumn{2}{c||}{$\bt = 10$} & \multicolumn{2}{c|}{$\bt = 50$} \\
 \cline{2-9}
 & $\mu_{g, \rm kin}^h$& $\mu_g^h$ &$\mu_{g, \rm kin}^h$ & $\mu_g^h$& $\mu_{g, \rm kin}^h$ 
 & $\mu_g^h$&$\mu_{g, \rm kin}^h$ & $\mu_g^h$\\
 \hline
0.3 &  0.9625 &1.4860  & 0.9834 & 3.5006 & 0.9912 & 6.0028 &  1.0002 & 26.006  \\
0.5 & 0.9806 & 1.5291 &1.0211 & 3.5861 & 1.0463 &  6.1037 & 1.0949  & 26.125  \\
0.7 & 1.0302 & 1.6052 & 1.0860 & 3.7289 & 1.1325 & 6.2862 & 1.2619  & 26.393  \\
0.9 & 1.1122 & 1.7138 & 1.1783 & 3.9224 & 1.2454 & 6.5442 & 1.4862  & 26.846  \\
1 & 1.1662 & 1.7813& 1.2359& 4.0373& 1.3124& 6.7001&1.6158 &27.150\\
1.1 & 1.2299 & 1.8582 & 1.3020 &  4.1647& 1.3873 & 6.8736 & 1.7562 & 27.506  \\
1.3 & 1.3902 & 2.0445 & 1.4638 & 4.4586 & 1.5641 & 7.2743 & 2.0690 & 28.375 \\
1.5 & 1.6029 & 2.2824 & 1.6741 & 4.8106 & 1.7850 & 7.7506 & 2.4280 & 29.455 \\
1.7 & 1.8822 & 2.5858 & 1.9477 & 5.2312 & 2.0637 & 8.3110 & 2.8422 & 30.754 \\
1.9 &  2.2474& 2.9743 & 2.3050  & 5.7366 & 2.4199 & 8.9684 & 3.3249 & 32.284  \\
 \hline
\end{tabular}
\caption{The simulated eigenvalue $\mu_g^h$  and the kinetic energy $\mu_{g,\rm kin}^h$ of the 
ground states of the fractional nonlinear Schr\"{o}dinger equation in an infinite potential well. 
Both the eigenvalue $\mu_g$ and the kinetic energy $\mu_g$ increases as $\ap$ or $\bt$ increases. 
If $\bt \gg 1$, the eigenvalue $\mu_g \sim O(\bt/2)$.} \label{T3}
\end{center}\vskip -10pt
\end{table}
%******************************************************************
Table \ref{T3} shows that both the eigenvalue $\mu_g$ and the kinetic energy $\mu_{g, \rm kin}$ 
increase as the parameter $\ap$ or $\bt$ increases.  
Comparing Tables \ref{T1} and \ref{T3}, we find that for a fixed $\ap$, the kinetic 
energy $\mu_{g,\rm kin}$ increases as the local interaction parameter $\bt$ increases.
When $\bt$ is small (e.g., $\bt = 1$), the kinetic energy is significant in the eigenvalue, 
while $\bt$ is large (e.g., $\bt = 50$), the local interactions become strong, and as 
a result the energy from the local interactions becomes significant.
Furthermore if $\bt \gg 1$, the eigenvalue $\mu_g \sim O(\bt/2)$.

% =============================================================================
\subsection{The first excited states}
\label{section5-2}

%%%% %%%% %%%% %%%% P1:
Figure \ref{F7} shows the wave function $\phi_1(x)$ of the  first excited states of 
%  where the arrow indicates the change 
%in wave functions for progressively increasing $\ap$.  
%******************************************************************
\begin{figure}[htb!]
\centerline{
\includegraphics[height=4.8600cm,width=6.660cm]{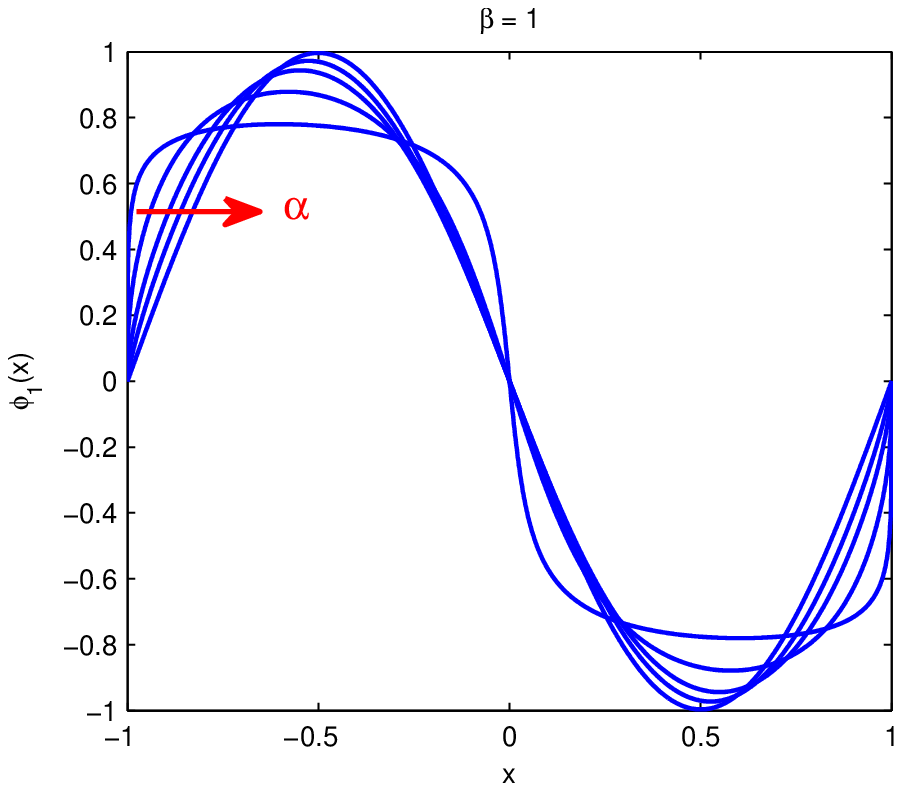}
\includegraphics[height=4.8600cm,width=6.660cm]{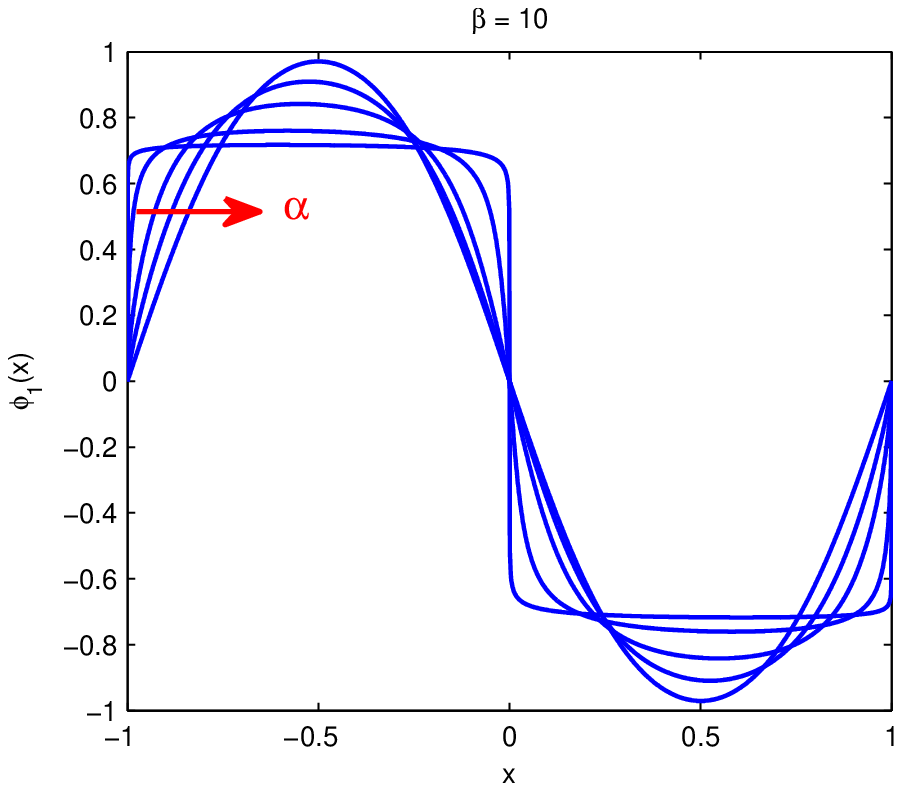}
}
\centerline{
\includegraphics[height=4.8600cm,width=6.660cm]{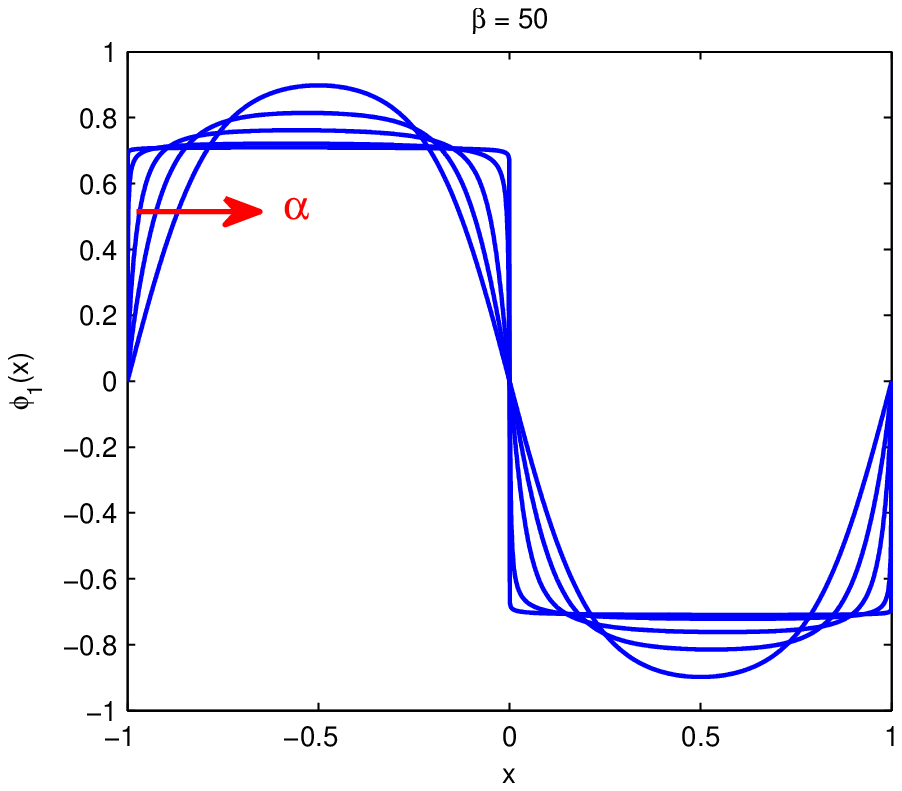}
\includegraphics[height=4.8600cm,width=6.660cm]{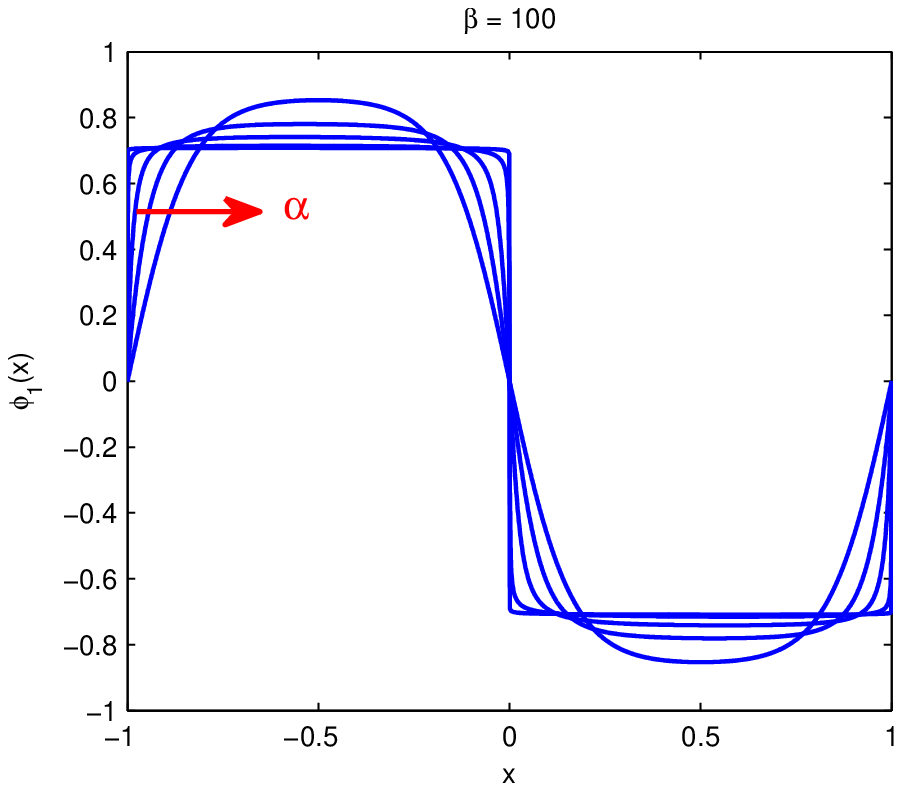}
}
\caption{The first excited states of the fractional nonlinear  Schr\"{o}dinger 
equation for $\ap = 0.2, 0.7, 1.1, 1.5$, and $1.99$, where the arrow 
indicates then change in the wave function for progressively increasing $\ap$.
When $\ap$ is small or $\bt$ is large, boundary layers exist at $x = \pm 1$ and  one inner layer 
emerges at $x = 0$.}\label{F7}\vskip -10pt
\end{figure}
%******************************************************************
the fractional nonlinear  Schr\"{o}dinger equation in an infinite  potential well.
The wave function $\phi_1(x)$ is antisymmetric with respect to the center of the potential well $x = 0$, 
independent of the parameters $\ap$ and $\bt$. As $\ap \to 2$, the wave function
converges to the first excited state solution of the standard Schr\"{o}dinger equation
 with the same nonlinear  parameter $\bt$.
The effect of local interactions becomes more significant when $\ap$ is small, resulting in 
sharp boundary layers at  $x = \pm 1$ as well as one inner layer at $x = 0$. 
The width of the boundary and inner layers decreases as $\ap$ decreases or $\bt$ increases. 
Numerical simulations show that our method converges fast in computing both the 
ground and first excited states.

%%%% %%%% %%%% %%%%
In addition,  Figure \ref{F12} shows the variance in position of the first excited states of the
 fractional nonlinear Schr\"{o}dinger equation for different parameters $\ap$ and $\bt$. 
%******************************************************************
\begin{figure}[ht!]
\centerline{
\includegraphics[height=5.06cm,width=6.960cm]{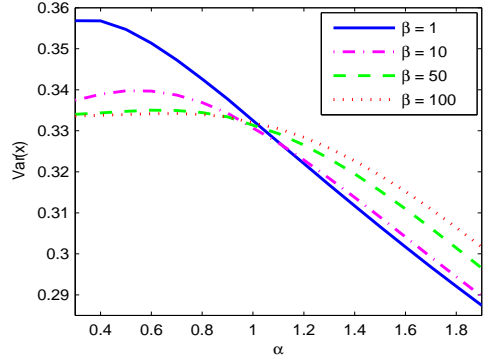}}
\caption{The variance in position of the first excited states of the fractional nonlinear Schr\"{o}dinger 
equation in an infinite potential well. It increases as the parameter $\ap$ 
decreases. The variance converges to $1/3$ as $\ap \to 0$ and $\bt \to \infty$. }\label{F12}\vskip -10pt
\end{figure}
%******************************************************************
Note that the expected value of position $\langle x \rangle_1\equiv 0$, independent of 
the parameters $\ap$ and $\bt$.  The properties of the variance in 
position can be divided into two cases:  when $\ap < 1$, 
the larger the parameter $\bt$, the smaller the variance. As $\bt \to \infty$, the variance 
converges to $1/3$. When $\ap > 1$, the larger the parameter $\bt$, the bigger the 
variance, implying that in this case the scattering of particles is mainly contributed  by  
the local repulsive interactions.

%%%% %%%% %%%% %%%%
In Table \ref{T4}, we present the numerical results of the eigenvalue $\mu_1^h$ and the 
kinetic energy $\mu_{1, \rm kin}$ of the first excited states for 
various  $\ap$ and $\bt$.  
%******************************************************************
\begin{table}[h!]
\begin{center}
\begin{tabular}{|c||c|c||c|c||c|c||c|c|}
\hline
\multirow{2}{*}{$\ap$} &  \multicolumn{2}{c||}{$\bt = 1$} &  \multicolumn{2}{c||}{$\bt = 5$} &  \multicolumn{2}{c||}{$\bt = 10$} & 
 \multicolumn{2}{c|}{$\bt = 50$} \\
 \cline{2-9}
 & $\mu_{1, \rm kin}^h$& $\mu_1^h$ &$\mu_{1, \rm kin}^h$ & $\mu_1^h$& $\mu_{1, \rm kin}^h$ 
 & $\mu_1^h$&$\mu_{1, \rm kin}^h$ & $\mu_1^h$\\
 \hline
0.3 & 1.3347 & 1.8994 &1.3919  &  3.9536& 1.4193 & 6.4629 &1.4544 & 26.471   \\
0.5 & 1.6176 & 2.2248 & 1.7011 & 4.3873 & 1.7670&  6.9514&1.9258 & 27.036  \\
0.7 & 1.9850 &2.6253  & 2.0763 & 4.9161 & 2.1727&  7.5807&2.5141 & 27.932    \\
0.9 & 2.4665 & 3.1325 & 2.5532 & 5.5554 & 2.6681& 8.3519 &3.2042 & 29.195   \\
1 & 2.7616 & 3.4385& 2.8430& 5.9257& 2.9622& 8.7962&3.5912 &29.968\\
1.1 & 3.1009 & 3.7874 & 3.1754 & 6.3365 & 3.2960& 9.2847 &4.0113 &30.835     \\
1.3 &  3.9416 & 4.6449 & 4.0007 & 7.3079 & 4.1145 & 10.417 &4.9744 & 32.866    \\
1.5 & 5.0624 &  5.7798& 5.1062 & 8.5416 & 5.2040 & 11.809  &6.1553 & 35.321   \\
1.7 & 6.5661&  7.2961& 6.5970 &  10.141& 6.6742 & 13.554 &7.6466 & 38.255   \\
1.9 & 8.5968 & 9.3383 & 8.6180 &  12.254& 8.6749 & 15.793 &9.5898 & 41.752    \\
 \hline
\end{tabular}
\caption{The simulated eigenvalue $\mu_1^h$  and the kinetic energy $\mu_{1,\rm kin}^h$ of the first excited 
states of the fractional nonlinear Schr\"{o}dinger equation in an infinite potential well. Both the eigenvalue 
and the kinetic energy increases as $\ap$ or $\bt$ increases. } \label{T4}
\end{center}
\end{table}
It shows that both the eigenvalue and  kinetic energy increase as the parameter $\ap$ or $\bt$ 
increases. Comparing Tables \ref{T4} and \ref{T3}, we find that for fixed parameters $\ap$ and $\bt$, 
 the kinetic energy $\mu_{1, \rm kin}$ of the first excited states is much larger than 
 $\mu_{g, \rm kin}$ of the ground states, which mainly caused by the emergence of the  
 inner layer in the first excited states.  Similar to the cases of the ground states, the kinetic  
 energy is significant in the eigenvalue only when the local interaction is weak (i.e., $\bt$ is small).

%******************************************************************
% ==========================================================================
\section{Conclusion and discussion}
\label{section6}

%%%% %%%% %%%% %%%% 
Due to the non-locality  of the fractional Laplacian, it is very challenging to solve 
the eigenvalues and eigenfunctions of the fractional Schr\"{o}dinger equation in an 
infinite potential well. We proposed an efficient and accurate numerical method to compute the 
ground and first excited states of the 
one-dimensional (1D) fractional Schr\"{o}dinger equation  in an infinite potential well, 
based on the integral representation of the fractional Laplacian $-(-\Dt)^{\ap/2}$. 
%In the literature \cite{Laskin2000, Laskin2000-1, Laskin2000-2, Dong2007, 
%Yang2010, Bayin2012, Bayin2013},  the eigenvalues and eigenfunctions of the 
%fractional Schr\"{o}dinger equation in an infinite potential well are highly debated based 
%on the pseudo-differential representation of the fractional Laplacian $-(-\Dt)^{\ap/2}$, 
%however, so far no agreement has been reached on this problem. 
%is usually treated equivalently as the fractional Schr\"{o}dinger equation in a bounded domain with 
%two-point  homogeneous Dirichlet boundary conditions.
%However, the Riesz fractional Laplacian is a nonlocal operator, and the wave functions 
%interact globally in the whole space ${\mathbb R}$.  Hence,  it is wrong to consider the boundary 
%conditions only at two points.
Taking the nonlocal character of the fractional Laplacian into account, we  introduced the 
nonzero volume constraint to the fractional Schr\"{o}dinger equation in a bounded domain. 
To compute its ground and first excited states,  we proposed a fractional gradient flow with 
discrete normalization and discretized it by
the trapezoidal type quadrature rule method in space and the semi-implicit Euler method in time.
Our method can be used to compute the ground and first excited states for both 
linear and nonlinear fractional Schr\"{o}dinger equation. 
It can be also applied to study the fractional partial differential equations (PDEs) with Riesz 
 fractional derivatives in space.

%%%% %%%% %%%% %%%%
To demonstrate the effectiveness of our numerical method,   
on the one hand, we studied the ground and first excited states in the linear 
($\bt = 0$) cases. Our numerical results suggested that the eigenfunctions of the fractional 
Schr\"{o}dinger equation
 are significantly different from those of the standard Schr\"{o}dinger equations. 
%and thus the results in the literature  
%\cite{Laskin2000, Laskin2000-1, Laskin2000-2, Dong2007, Bayin2012, Bayin2013, Dong2013}
%are incorrect.  
The nonlocal interactions are strong when $\ap$ is small, 
leading to large scattering of particles in the infinite potential well. 
Compared to the ground states, the scattering of particles in the first excited states is stronger 
because of its larger energy.
In addition, we compared our simulated eigenvalues with the 
estimates obtained in \cite{Chen2005, Deblassie2004, Banuelos2004, Kwasnicki2012} and 
found that the asymptotic approximations obtained   in 
\cite{Kwasnicki2012} are more accurate when $\ap \to 2$. 
The estimates of lower and upper bounds of the eigenvalues  in \cite{Chen2005, Deblassie2004} are 
valid for both ground and first excited states, however,  the estimates obtained in \cite{Banuelos2004} 
for the ground states are much sharper. 

%%%% %%%% %%%% %%%%
On the other hand,  we studied the ground and first excited states in the nonlinear cases with 
$\bt > 0$.  It showed that the eigenvalues of both ground and first excited states 
monotonically  increase as the parameter $\ap$ or $\bt$ increases. However, 
the presence of the local interactions in the fractional Schr\"{o}dinger equation significantly 
changes the wave functions, especially when $\ap$ is small.  
Boundary layers emerge in the ground states, and additionally one inner layer exists in the first 
excited states. The width of the boundary layers depends on both the local interaction 
parameter $\bt$ and the nonlocal interaction parameter $\ap$. 
More precisely, if $\bt$ is fixed,  the width of the boundary layers increases as  a 
power function of $\ap$; while if $\ap$ is fixed,  the width decreases as a power function of  $\bt$.   
For the standard nonlinear Schr\"{o}dinger equation, the width of the boundary layers 
is $O(1/\sqrt{\bt})$ \cite{Bao2007}.
In addition,  the expected value of position is always zero, independent of the parameters 
$\ap$ and $\bt$. However,  the variance in position increases as $\ap$ decreases or $\bt$ 
increases,  implying that strong local or nonlocal interactions result in the large scattering of particles.
As $\ap \to 0$ or $\bt \to \infty$,  the variance in position convergences to $1/3$.

\bigskip 

\noindent {\bf Acknowledgements }  The authors would like to thank Dr. John Burkardt and Dr. 
Qiang Du for their valuable comments to improve the paper. 
% ==========================================================================

\end{document}